\documentclass[11pt,a4paper]{amsart}
\usepackage[a4paper, margin=3cm]{geometry}
\newtheorem{theorem}{Theorem}

\usepackage{graphicx}
\usepackage{amsmath}
\usepackage{amssymb}
\usepackage{rotating}
\usepackage{mathtools}

\usepackage{url}
\mathtoolsset{showonlyrefs}

\usepackage[utf8]{inputenc}
\usepackage{chicago}
\usepackage{latexsym}
\usepackage{xcolor}
\usepackage{tabularx} 
\usepackage{booktabs}
\usepackage{adjustbox}
\usepackage{framed}

\usepackage{environ}
\usepackage{multirow}
\usepackage{caption}

\newcommand{\eb}{EB}
\newcommand{\est}[2]{\hat{\beta}^{#1}_{\mbox{\scriptsize #2}}}
\newcommand{\esto}[1]{\hat{\beta}_{\mbox{\scriptsize #1}}}
 
\usepackage{pdflscape}%For orizontal table
\usepackage{lipsum}%To set new counter for the examples
\usepackage{enumitem}%List with roman numbers (List of papers)

\newcounter{assumptionc}
\newcounter{corollaryc}
\newcounter{lemmac}
\newcounter{remarkc}
\newcounter{examplec}
\newcounter{propositionc}

\newtheorem{assumption}[assumptionc]{Assumption}
\newtheorem{corollary}[corollaryc]{Corollary}
\newtheorem{lemma}[lemmac]{Lemma}
\newtheorem{proposition}[propositionc]{Proposition}
\newtheorem{remark}[remarkc]{Remark}
\newtheorem{example}[examplec]{Example}

\newcommand{\kl}{KL}
\newcommand{\qd}{QR}

\newcommand{\indep}{\perp\hskip -7pt \perp }

\newcommand{\caps}[1]{Simulation results for sample size $N=#1$, based on 1000 replications.
Bias (BIAS), standard deviation (SD), and root mean square error (RMSE).} 

\newcommand{\capS}[1]{Simulation results for sample size $N=#1$, based on 1000 replications.
Empirical coverage of $95\%$ confidence intervals using normal approximation, $\hat{\beta} \pm 1.96\sqrt{ \hat{\tau}^2/N} $,
empirical standard errors ($\text{SD}$), and mean of estimated standard errors ($\hat{\tau}$).
} 

\makeatletter
\newcommand{\addresseshere}{
	\enddoc@text\let\enddoc@text\relax
}
\makeatother

\begin{document}
\numberwithin{equation}{section}

\title[Entropy balancing estimation]{Large sample properties of entropy balancing estimators of average causal effects}
\author{David K\"{a}llberg\textsuperscript{1}}
\author{Ingeborg Waernbaum\textsuperscript{2,3}}

\address{
\textsuperscript{1} Department of Statistics, USBE, Umeå University, SE-901 87, Umeå, Sweden\\
\textsuperscript{2} Department of Statistics, Uppsala University, Box 513, SE-751 20, Uppsala, Sweden\\
\textsuperscript{3} Institute for Evaluation of Labour Market and Education Policy (IFAU), Box 513, SE-751 20, Uppsala, Sweden}
\email{david.kallberg@umu.se}

\begin{abstract}
Weighting methods are used in observational studies to adjust for covariate imbalances between treatment and control groups. Entropy balancing (EB) is an alternative to inverse probability weighting with an estimated propensity score. The EB weights are constructed to satisfy balance constraints and optimized towards stability.
We describe large sample properties of EB estimators of the average causal treatment effect, based on the Kullback-Leibler and quadratic R\'{e}nyi relative entropies. Additionally, we propose estimators of their asymptotic variances. Even though the objective of EB is to reduce model dependence, the estimators are generally not consistent unless implicit parametric assumptions for the propensity score or conditional outcomes are met. The finite sample properties of the estimators are investigated through a simulation study. In an application with observational data from the Swedish Childhood Diabetes Register, we estimate the average effect of school achievements on hospitalization due to acute complications of type 1 diabetes mellitus.

\end{abstract}

\maketitle


\addresseshere 

\newpage

\section{Introduction}
\label{s:intro}

When estimating a causal effect of a treatment with observational data, methods that control for confounding using propensity scores \cite{RR:83} are extensively used.
 The propensity score, the conditional probability of receiving a specific treatment conditional on observed covariates, are used in estimators for average causal effects in several different approaches, such as weighting, matching and regression adjustment, see e.g, Goetghebeur et al (2020) \nocite{goetghebeur2020formulating} 
 for an overview. 
 In particular, inverse probability weighting (IPW) plays a central role in the theory of semiparametric estimation with missing data \cite{AT:07,seaman2013review}. 
 A prototypical IPW estimator utilises a parametric model for the propensity score \cite{LD:04,il2019efficiency}. 
 The parametric IPW estimator relies on the correct specification of the propensity score model although IPW estimators under various semi-parametric assumptions of the propensity score have also been proposed \cite{HIR:03,van2014targeted,liu2018alternative}. 
 
 As an extension to model-based IPW, a class of estimators referred to as calibration or balancing estimators has been developed \cite{hainmueller2012entropy,chan2014oracle,chan2016globally,tan2020regularized}. 
In this class, the observed sample is reweighted by minimizing a distance measure or a loss function while imposing a set of balance constraints for the treated and controls in the reweighted sample in terms of means of functions of the confounding covariates. A similar procedure is used by the calibration estimators in unequal probability sampling \cite{deville1992calibration}.
As a hybrid between IPW and balancing, \citeN{IR:14} use a combination of parametric propensity score modeling and calibration by adding estimating equations with balancing constraints for the covariates to the score equations of the propensity score model. Properties of an extended version of the estimation approach targeting a high-dimensional setting is derived in \citeN{ning2020robust}. 
These approaches relate to the general framework introduced in \citeN{zhao2019covariate} in which the propensity score is estimated using loss functions that are designed to give covariate balance. In a recent study, \citeN{chattopadhyay2020balancing}, referring broadly to balancing vs modelling, review and compare the two weighting approaches. 

Several studies have considered weighting procedures derived from balance criteria beyond exact equality of sample moments.
\citeN{zubizarreta2015stable} studies minimum-variance weights under approximate equality of moments and derives a link between the pre-specified level of tolerated imbalance and estimator bias given weak nonparametric assumptions for the outcome (response surface). For the same class of estimators,  \citeN{wang2020minimal} presents further developments derived by a connection between approximate covariate balance and shrinkage estimation of the inverse propensity score.
\citeN{yiu2018covariate} also consider minimum-variance weights, but under a balancing constraint derived from a propensity score model, which imposes weights to eliminate the association between covariates and treatment in the sample.
\citeN{wong2017kernel} derive asymptotic properties of a weighting estimator obtained by optimizing balance over an infinite-dimensional reproducing-kernel Hilbert space of covariate functions.  

In this paper we derive asymptotic properties for two prototypical entropy balancing estimators of the average treatment effect (ATE). Entropy balancing estimators uses balancing weights that minimize the divergence, or relative entropy, relative a set of uniform weights. By construction, the weights mean balance a set of covariate functions between the treated, controls, and the combined group, a property referred to as three-way balance.  Two variants of the relative entropy are considered: the Kullback-Leibler divergence (\kl), initially proposed by \citeN{hainmueller2012entropy} (for the average treatment effect of the treated, ATT), and the quadratic R\'{e}nyi divergence (\qd), 
which yields the minimum variance weights. Entropy balancing does not explicitly use models for the treatment or the outcome, a distinctive feature compared to other three-way balancing estimators, for example the calibration estimator in \citeN{tan2020regularized} and the empirical likelihood approach \cite{qin2007empirical} which both utilize a parametric model for the propensity score.

Although reducing model dependence was expressed by \citeN{hainmueller2012entropy} as a main motivation to the introduction of entropy balancing, estimators obtained by minimizing the two divergences (henceforth referred to as {\kl} and {\qd}  estimators) entail implicit functional forms given by the two divergence measures respectively. This means that the correctness of the implied propensity score models plays a role for the consistency of the estimators. 
Under additional assumptions of models for the outcome, we show that the {\kl} and {\qd} estimators are consistent for the average causal effect and asymptotically normal.  The outcome model assumption can be relaxed by separating and combining assumptions for the propensity score and the outcome for the treated and controls. A notable difference for the role of the outcome model assumptions between the {\kl} and {\qd} estimators compared to, e.g., regression imputation estimators or augmented IPW-estimators is that the regression model per se is not used in the estimators. Instead, outcome regression model assumptions are (implicitly) included in the balancing constraints selected in the divergence minimization scheme. Additionally, we propose an estimator for the variance. The entropy balancing estimators and the proposed variance estimator are evaluated in a simulation study. The estimators are applied to a motivating example on the effect of school achievements on acute complications of childhood onset type 1 diabetes mellitus.

The entropy balancing estimators under fixed balance constraints studied here can be seen as parametric versions of the more general nonparametric calibration estimators described in  \citeN{chan2016globally} and \citeN{wang2020minimal} where the number of balance constraints are assumed to grow with the sample size. Although these nonparametric versions constitute a larger class of asymptotically optimal calibration estimators, our approach provides guidance on the magnitude of the error resulting from the use of prototypical versions of the estimators applied in practice. 

The remaining of the paper is organized as follows. 
Section \ref{sec2} contains notation and assumptions together with a brief background to the problem.
In Section~\ref{sec3}, we introduce entropy balancing and provide details on the two particular cases that are studied here: the Kullback-Leibler- and quadratic R\'{e}nyi estimators.
In Section~\ref{sec4}, we state our main results, including asymptotic limits for the estimation error, and conditions for consistency and asymptotic normality. 
Section~\ref{sec5} presents a simulation study that evaluates the estimators on an extension of a design previously used in the literature. A data example, with a study of acute complications of type 1 diabetes mellitus is provided in Section~\ref{sec6} and in Section~\ref{sec7} we conclude with a discussion. 

\section{Model and Theory}\label{sec2}
We use the potential outcome framework \cite{Neyman23,DR:74} to define a causal effect of a treatment on an outcome of interest. Consider a binary treatment, $T=1$ for individuals under treatment and $T=0$ for individuals under no treatment. We define $Y(1)$ as the potential outcome under treatment and $Y(0)$ as the potential outcome under the control treatment and assume that the observed $Y=TY(1)+(1-T)Y(0)$. We let $X$ denote a $p$-dimensional vector of covariates, $X=(X_1,\ldots,X_j,\ldots X_p)$. Thus, for each unit $i$ in an iid sample, $i=1,\ldots N$, we observe the vector $(T_i,X_i,Y_i)$ and for simplicity we omit the index $i$ when not needed. The probability of treatment conditional on the covariates, the propensity score, is denoted by $e(X)=P(T=1|X)$. 
We consider the average causal effect $\beta=E\left[Y(1)-Y(0)\right]$ as the estimand of interest, and let $\beta_1 = E[Y(1)]$ and $\beta_0=E[Y(0)]$ be the marginal potential outcome means. The conditional mean functions are denoted by $\beta_1(X) = E[Y(1)|X]$ and $\beta_0(X) = E[Y(0)|X]$. 
We use the notation $\stackrel{wp1}{\to}$ for almost sure convergence (with probability 1), and let $\stackrel{P}{\to}$ and $\stackrel{D}{\to}$ denote convergence in probability and in distribution, respectively.

For identification of $\beta$ with the observed data we use the assumption of no unmeasured confounding and overlap as defined below.

\begin{assumption}[No unmeasured confounding]\label{nuc}
	$Y(t)\indep T\mid X, t=0,1.$
\end{assumption}

\begin{assumption}[Overlap]\label{overlap}
	$\eta<P(T=1|X)<1-\eta$, for some $\eta>0$.
\end{assumption}
Under Assumptions \ref{nuc} and \ref{overlap} the average causal effect can be identified with the observed data since
\begin{equation}\label{id}
\beta = E\left[\frac{TY}{e(X)}\right]-E\left[\frac{(1-T)Y}{1-e(X)}\right]  
\end{equation}
In the conventional IPW approach, an empirical version of (\ref{id}) is used to estimate $\beta$, such as
\begin{equation}\label{ipw} 
\hat{\beta}_{IPW} =  \frac{1}{N}   \sum_{i=1}^N w_i T_i Y_i - \frac{1}{N} \sum_{i=1}^N w_i (1-T_i) Y_i,
\end{equation}
and $w_i = T_i \hat{e}(X_i)^{-1} + (1-T_i) (1 - \hat{e}(X_i))^{-1}$, where $\hat{e}(X)$ is an estimate of the propensity score $e(X)$.
The missing outcomes for the respective groups are generated with $w$ and the estimator $\hat{\beta}_{IPW}$ is obtained as the difference in mean outcomes between the groups in the generated data. 

Due to sensitivity to model misspecification and poor finite sample performance of conventional IPW estimators \cite{waernbaum2012model}, 
it has been suggested to derive weights $w_i$ on the basis of covariate balance rather than propensity score estimation. 
In this setting, weighting by the inverse propensity $e(X)^{-1}$ is used as a change of probability measure that balances the covariate distribution between the treatment group and the combined group, and likewise $(1-e(X))^{-1}$ creates covariate balance between the control group and the combined group. 
This is referred to as three-way balance in the sequel and is a formulation of the following property: 
for any real-valued function $u(X)$ with $E[u(X)] < \infty$,
\begin{equation}\label{balance} 
E\left[\frac{T u(X)}{e(X)}   \right] = E\left[\frac{(1-T) u(X)}{1-e(X)}   \right] = E[u(X)].
\end{equation}
Here, we consider a finite-dimensional version of \eqref{balance} which is specified by a set of user-defined functions $u_1(X),\ldots, u_K(X)$.
It is common to consider moment constraints up to some degree, for example,  
if we let $K$ equal the dimension $p$ of $X$, and put $u_k(X) = X_k$, first order mean balance is obtained. We use the notation $U_K(X) = (u_1(X),\ldots, u_K(X))$ and $M_K = (m_1,\ldots,m_K)$, 
where $m_k =  E[u_k(X)]$, and let $\hat{M}_K = (\hat{m}_1,\ldots,\hat{m}_K)$ denote the corresponding sample vector. 
The following regularity assumption for the balance functions is used throughout the paper. Similar assumptions are used in \citeN{HIR:03} and \citeN{chan2016globally}.
\begin{assumption} 
For some $C > 0$, $|u_k(X)| \leq C,k =1,\ldots,K,$ almost surely, and the random vector $U_K(X) = (u_1(X),\ldots,u_K(X))$ has a nonsingular covariance matrix. 
\label{as:Ux} 
\end{assumption}

\section{Entropy Balancing}\label{sec3}
Entropy balancing was originally proposed by \citeN{hainmueller2012entropy}.  
Although presented as a general weighting procedure for pre-processing observational data, 
the paper introduced the method by considering estimation of the average treatment effect of the treated, $E[Y(1)-Y(0)\mid T=1]$. 

Here we study entropy balancing for the problem of estimating the average treatment effect $\beta = E[Y(1) - Y(0)]$.
The estimator $\hat{\beta}$ is of the same form as the IPW-estimator \eqref{ipw}, but the weights $w_1,\ldots,w_N$ are derived on the basis of balancing property \eqref{balance} instead of propensity score estimation.
The weight vector $w=(w_1,\ldots,w_N)$ is obtained by minimizing the divergence $D(w||d)$, or relative entropy, between $w$ and a vector of reference weights $d=(d_1,\ldots,d_N)$ while imposing empirical mean balance with respect to the functions $u_1(X),\ldots, u_K(X)$. 
To be specific, $w$ is found as the solution of the optimization problem to find
\begin{equation}\label{Dw}
\displaystyle{
\min_w D(w || d)
}
\end{equation}
subject to the three-way balance- and normalizing constraints
\begin{equation} \label{eq:balance}
\frac{1}{N} \sum_{\{i:T_i=1\}} w_i u_k(X_i)   = \frac{1}{N} \sum_{\{i:T_i=0\}} w_i u_k(X_i) = \frac{1}{N} \sum_{i=1}^n u_k(X_i) = \hat{m}_k ,   \quad k = 1,\ldots,K,    
\end{equation}
\[
\sum_{\{i:T_i=1\}} w_i 
 =  \sum_{\{i:T_i=0\}} w_i   = N.
\]
The divergence measure $D(w || d)$ may be chosen in different ways. 
In information theory it is standard to apply a member of the {\it R\'{e}nyi divergences}, a parametric family defined by
\begin{equation}
D(w || d) = D_\alpha(w || d) = \frac{1}{\alpha-1} \log \left( \sum_{i=1} ^N \frac{w_i^\alpha}{d_i^{\alpha-1}} \right ), \quad \alpha > 0. 
\end{equation}
With a divergence from this class, problem \eqref{Dw} divides into two problems, one for each group, that can be solved separately. In this paper we investigate two of the most used R\'{e}nyi divergences: the Kullback-Leibler divergence ({\kl}) ($\alpha=1$) and the quadratic R\'{e}nyi divergence ({\qd}) ($\alpha= 2$), given by
\[
D_1(w || d)  = \sum_{i=1}^N w_i \log (w_i/d_i), 
\]
 \[
D_2(w || d)  = \log \left( \sum_{i=1}^N w_i^2/d_i \right).
 \]

The base weights $d_1,\ldots,d_N$ are considered as fixed and can represent some prior information although a common choice are uniform base weights, i.e., $d_i=1$ \cite{hainmueller2012entropy,chan2016globally}.  
In this study we use uniform base weights, which entails that the entropy of the weights is maximized \cite{kallberg2012statistical}. Next we provide details on how to obtain entropy balancing estimators using KL- and QR divergences, respectively. 

\subsection{Kullback-Leibler divergence}
When the KL-divergence is used, the entropy balancing weights $\{ w^{\mbox{\scriptsize KL}}_{t,i} : T_i = t\}$ for group $t =1,0$, is obtained by solving
\begin{equation}\label{obj1}
\min \sum_{\{i:T_i = t\}}  w_i\log w_i
\end{equation}
subject to the constraints 
\begin{equation} \label{eq:balance2} 
 \frac{1}{N} \sum_{\{i:T_i=t\}} w_i u_k(X_i)  = \hat{m}_k,  
 \end{equation}
 \[
 \sum_{\{i:T_i=t\}} w_i   = N. 
\]
The solution of this convex and linearly constrained problem can be represented by an {\it exponential tilting} \cite{erlander1981entropy}, given by
\begin{equation}\label{eq:expTilting}
w^{\mbox{\scriptsize KL}}_{t,i} = N \frac{ \exp ( -\sum_{k=1}^K \hat{\lambda}_{t,k} u_k(X_i) ) } { \sum_{T_i = t}  \exp ( -\sum_{k=1}^K \hat{\lambda}_{t,k} u_k(X_i) )}, \quad T_i = t ,
\end{equation}
where the $K$-vector $\hat{\lambda}_t = (\hat{\lambda}_{1,t}, \ldots,\hat{\lambda}_{K,t})$ solves the unconstrained dual problem
\begin{equation}\label{dual1}
\min_{\lambda_t \in R^K} g_t(\lambda_t), \quad g_t(\lambda_t) = 
\log\left(\sum_{\{i:T_i=t\}}  \exp\left(-\sum_{k=1}^K \lambda_{t,k} u_k(X_i) \right) \right) + \sum_{k=1}^K \lambda_{t,k}  \hat{m}_k.
\end{equation}
The entropy balancing estimator corresponding to this weighting is given by
\begin{equation}\label{est:kl}
\esto{\kl} =
 \frac{1}{N} 
 \sum_{\{i:T_i=1\}} w^{\mbox{\scriptsize{\kl}}}_{1,i}  Y_i - 
 \frac{1}{N} 
 \sum_{\{i:T_i=0\}} w^{\mbox{\scriptsize{\kl}}}_{0,i}  Y_i.
\end{equation}
Problem \eqref{dual1} can be solved by applying an iterative Newton-type algorithm, e.g., as described in \citeN{hainmueller2012entropy}. This requires that the solution $\hat{\lambda}_t$ is unique, which holds when the observed covariate vectors $\{X_i : T_i = t\}$ spans $R^K$, making the objective function $g_t(\cdot)$ strictly convex. 
We set $\hat{\lambda}_t = 0^K = (0,0, \ldots, 0)$ when a unique solution of \eqref{dual1} does not exist, in which case $\hat{\beta}{\mbox{\scriptsize KL}}$ becomes the (unadjusted) difference in mean. 
More intricate alternatives to this extended definition are possible, but since it is only technical and serves to enable asymptotic theory for the estimators, our crude variant is sufficient for our purposes. 

\subsection{Quadratic R\'{e}nyi divergence} 
The entropy balancing weights for the QR-divergence in group $t =1,0$, denoted by $\{w^{\mbox{\scriptsize \qd}}_{t,i} : T_i = t\}$, are found as the solution of
\begin{equation}\label{obj:QR}
\min \sum_{\{i:T_i = t\}}  w_i^2
\end{equation}
subject to balancing condition \eqref{eq:balance2}. 
This is a quadratic problem with linear constraints, and consequently has a straightforward solution. 
Let $A_1$ and $A_0$ be $N \times (K+1)$ matrices where row $i$ of $A_1$ is given by $(T_i, T_i U_K(X_i))$, and row $i$ of $A_1$ is given by $(1-T_i, (1-T_i)U_K(X_i)), i=1,\ldots, N$. If $A_t$ has full rank, problem \eqref{obj:QR} admits a unique solution which can be represented by the linear expression
\begin{equation}\label{wqd}
w^{\mbox{\scriptsize{QR}}}_{t,i} = \hat{\kappa}_{t,0} + \sum_{k=1}^K\hat{\kappa}_{t,k}u_k(X_i), \quad T_i = t,
\end{equation}
where the $(K+1)$-vector $\hat{\kappa}_t = (\hat{\kappa}_{t,0}, \ldots,\hat{\kappa}_{t,K})$ is obtained as
\begin{equation}\label{dual2}
\hat{\kappa}_t  = (N^{-1}A_t^T A_t)^{-1} \hat{M}^*_{K}, \quad \hat{M}^*_{K} = (1, \hat{m}_1,\ldots, \hat{m}_K).
\end{equation}
The quadratic entropy balancing estimator is defined as
\begin{equation}\label{est:qd}
\esto{\qd} =
 \frac{1}{N}   \sum_{\{i:T_i=1\}} w^{\mbox{\scriptsize \qd}}_{1,i} Y_i - \frac{1}{N} \sum_{\{i: T_i=0\}} w^{\mbox{\scriptsize{\qd}}}_{0,i}  Y_i.
\end{equation}
Similarly as for the KL-estimator, we define $\hat{\kappa}_t = (N/N_t,0,\ldots, 0)$ in case $A_t$ does not have full rank, where $N_t$ denotes sample size of group $t=1,0$, which corresponds to the unadjusted difference in mean.\medskip

\section{Asymptotic properties of the KL and QR estimators}\label{sec4}
The previous section describes how the entropy balancing weights are parameterized by the solutions $\{\hat{\lambda}_1,\hat{\lambda}_0\}$ and $\{\hat{\kappa}_1,\hat{\kappa}_0\}$ of the dual problems \eqref{dual1} and \eqref{dual2}, respectively. 
The following result on convergence of the dual parameters is used to obtain asymptotic results for the corresponding estimators. Proofs of the results in this section are in Appendix A.
\begin{proposition} \label{prop:1} 
{\it Let Assumptions \ref{nuc}-\ref{as:Ux} be satisfied. Then, as $N \to \infty$,
\begin{itemize}
\item[(i)] $\hat{\lambda}_t \stackrel{wp1}{\to} \tilde{\lambda}_t$, where $\tilde{\lambda}_t = (\tilde{\lambda}_{t,1}, \ldots,\tilde{\lambda}_{t,K})$ is the $K$-vector that uniquely  solves the system
\[
E[e(X)^t(1-e(X))^{1-t}e^{(\tilde{\lambda}_t)^T U_K(X)}(u_k(X)-m_k)], \quad k=1,\ldots, K.
\]
\item[(ii)] $\hat{\kappa}_t \stackrel{wp1}{\to} \tilde{\kappa}_t$,  
where $\tilde{\kappa}_t = (\tilde{\kappa}_{t,0}, \ldots,\tilde{\kappa}_{t,K})$ is the $(K+1)$-vector that uniquely solves the (linear) system
\[
\begin{array}{l}
E[e(X)^t(1-e(X))^{1-t}    (\hat{\kappa}_t)^T(1, U_K(X))		u_k(X)] = m_k, \quad k=1,\ldots, K,  \\
E[e(X)^t(1-e(X))^{1-t}     (\hat{\kappa}_t)^T(1, U_K(X)) ] = 1.
\end{array}
\]
\end{itemize}
}
\end{proposition}
 
As described in Proposition \ref{prop:1}, the sequences $\{\hat{\lambda}_t\}$ and $\{\hat{\kappa}_t\}$ converge to certain constant vectors $\tilde{\lambda}_t$ and $\tilde{\kappa}_t$. Based on these limits, we introduce asymptotic counterparts of the entropy balancing weights.
For the KL estimator, let
\begin{equation}\label{limWkl}
\tilde{w}^{\mbox{\scriptsize \kl}}_t(X)  = \tilde{w}^{\mbox{\scriptsize \kl}}_t(X; \tilde{\lambda}_t) = \frac{e^{(\tilde{\lambda}_t)^T U_K(X)}}{E[e(X)^t (1-e(X))^{1-t} e^{(\tilde{\lambda}_t)^T U_K(X)}]}, \quad t = 1,0,
\end{equation}
and for the QR estimator, let
\begin{equation}\label{limWqd}
\tilde{w}^{\mbox{\scriptsize \qd}}_t(X) = \tilde{w}^{\mbox{\scriptsize \qd }}_t(X; \tilde{\kappa}_t)  = \tilde{\kappa}_{t,0} + \sum_{k=1}^K \tilde{\kappa}_{t,k} u_k(X), \quad t = 1,0.
\end{equation}
In the following theorem, these limiting weights are used to formulate a result on weak convergence for the entropy balancing estimators. The subsequent corollary gives representations of the asymptotic estimation error and for ease of presentation we use $\esto{\eb}$ to denote the KL and QR estimators for results that hold for both estimators. 
\begin{theorem} [Weak convergence] \label{th:1}  
If Assumptions \ref{nuc}-\ref{as:Ux} hold,  then
\begin{equation}
\esto{\eb} \stackrel{P}{\to}  
E[\tilde{w}^{\mbox{\scriptsize \eb }}_1(X) e(X) \beta_1(X)] - E[\tilde{w}^{\mbox{\scriptsize \eb} }_0(X) (1- e(X)) \beta_0(X)] \mbox{ as } N \to \infty.
\end{equation}
\end{theorem}
\begin{corollary}[Asymptotic error] \label{cor:1} If Assumptions \ref{nuc}-\ref{as:Ux} hold,  the asymptotic error satisfies
\[
\esto{\eb} - \beta \stackrel{P}{\to}  \tilde{\Delta}_{\mbox{\scriptsize \eb}} = 
Cov(Z^{\mbox{\scriptsize \eb}}_1(X),\beta_1(X)) - Cov(Z^{\mbox{\scriptsize \eb}}_0(X), \beta_0(X)) \mbox{ as } N \to \infty,
\]
where $Z^{\mbox{\scriptsize \eb}}_t(X) = \tilde{w}^{\mbox{\scriptsize \eb}}_t(X) e(X)^t (1-e(X))^{1-t}, t =1,0$. 
\end{corollary}
We see that the asymptotic error depends on the covariance of the propensity score model error and the conditional outcome. As indicated by Theorem \ref{th:1} and Corollary \ref{cor:1}, further assumptions are needed to ensure consistency of $\esto{\eb}$. As shown in the following corollary, it suffices to assume a linear outcome regression (OR) model, or a propensity score (PS) model combined with a linear OR model for one of the groups. The purpose of this result is to illustrate the idea and is not intended to be exhaustive, see Remark \ref{rem2}.
\begin{corollary}[Consistency, OR and PS models]\label{th:rob}
If Assumptions \ref{nuc}-\ref{as:Ux} hold, then for consistency $\esto{\kl} \stackrel{P}{\to} \beta$, each of the following conditions are sufficient:
\begin{itemize}
\item[(i)]  $\beta_1(X)$ and $\beta_0(X)$ are linear in $u_1(X),\ldots,u_K(X)$.  
\item[(ii)]  $\beta_0(X)$ and $\log[e(X)]$ are linear in $u_1(X),\ldots,u_K(X)$.  
\item[(iii)]  $\beta_1(X)$ and $\log[1-e(X)]$ are linear in $u_1(X),\ldots,u_K(X)$. 
\end{itemize}
A similar property hold for $\esto{\qd}$, with the PS models in (ii) and (iii) replaced by $e(X)^{-1}$ and $(1 -e(X))^{-1}$, respectively.
\end{corollary}
We note that a linear model assumption for $\beta_1(X)$ and $\beta_0(X)$ is not necessary for consistency of $\esto{\kl}$ ($\esto{\qd}$) but can be relaxed by imposing a log-linear (inverse-linear) functional form for the propensity score. 

\begin{remark}[Two-way balance]\label{rem2} 
The assumptions of Corollary \ref{th:rob} ensure that $\esto{\eb,1}$ and $\esto{\eb,0}$, the components of $\esto{\eb} = \esto{\eb,1} - \esto{\eb,0}$, are consistent as estimators of the group-wise parameters $\beta_t = E[Y(t)], t= 1,0$. This is sufficient but not necessary for consistency of $\hat{\beta}_{\mbox{\scriptsize \eb}}$.
For example, consider the case of no treatment effect, i.e., $\beta_1(X) = \beta_0(X) = \beta(X)$. Then the asymptotic estimation error is obtained from Theorem~\ref{th:1} as
$$
    \hat{\beta}_{\mbox{\scriptsize \eb}}   - 0
    \stackrel{P}{\to} 
    E[T\tilde{w}^{\mbox{\scriptsize  \eb}}_1(X) \beta(X)] - E[(1-T)\tilde{w}^{\mbox{\scriptsize \eb} }_0(X)  \beta(X)] \mbox{ as } N \to \infty. 
$$
Here $\hat{\beta}_{\mbox{\scriptsize \eb}}$ is consistent when the (asymptotic) weights two-way balance the common conditional mean $\beta(X)$ between the treatment and control group, i.e., when
$
E[T\tilde{w}^{\mbox{\scriptsize  }}_1(X) \beta(X)] = E[(1-T)\tilde{w}^{\mbox{\scriptsize } }_0(X)  \beta(X)]$.
In Appendix B we describe a class of distributions that results in this type of two-way balance.
\end{remark}
In the next theorem, we show that the linear model for $\beta_1(X)$ and $\beta_0(X)$, as in Corollary~\ref{th:rob}$(i)$, is sufficient to obtain asymptotic normality of the two studied estimators. 
\begin{theorem}[Asymptotic normality, OR model] \label{th:norm} If Assumptions \ref{nuc}-\ref{as:Ux} hold, and $\beta_1(X)$ and $\beta_0(X)$ are linear in $u_1(X),\ldots, u_K(X)$, then  
\begin{equation}
\sqrt{N} (\hat{\beta}_{\mbox{\scriptsize \eb}} - \beta) \stackrel{D}{\to} N(0, \tau^2_{{\mbox{\scriptsize \eb}}}) \mbox{ as } N \to \infty,
\end{equation}
where the asymptotic variance is given by
\[
\tau^2_{\mbox{\scriptsize \eb}}   =  
Var[\beta_1(X) - \beta_0(X)] +  E[ e(X) \tilde{w}^{\mbox{\scriptsize \eb}}_{1}(X)^2 \sigma^2_1(X) ] +  E[(1-e(X)) \tilde{w}^{\mbox{\scriptsize \eb}}_{0}(X)^2 \sigma^2_0(X)],
\]
and $\sigma^2_t(X) = Var[Y(t)|X]$. 
\end{theorem}
Note that differences in asymptotic variance between $\hat{\beta}_{\mbox{\scriptsize \kl}}$ and $\hat{\beta}_{\mbox{\scriptsize \qd}}$ are given by differences in the limiting weights. 
The large sample properties of the entropy balancing estimators described above are parametric versions of the calibration method studied by \citeN{chan2016globally}, where an estimator of $\beta$, denoted $\hat{\beta}_{\mbox{\scriptsize NP}}$, is obtained by three-way balancing a set of functions $u_1(X),\ldots,u_K(X)$ that grows with the sample size, i.e., $K=K(N) \to \infty$ as $N \to \infty$. Under nonparametric assumptions it is shown that $\sqrt{N}(\hat{\beta}_{\mbox{\scriptsize NP}} - \beta) \stackrel{D}{\to} N(0, \sigma^2_{eff})$, where $\sigma^2_{eff}$ is the semi-parametric efficiency bound \cite{Hahn1998}, given by
$
\sigma^2_{eff}   =  
Var[\beta_1(X) - \beta_0(X)] +  E[ e(X)^{-1}  \sigma^2_1(X) ] +  E[(1-e(X))^{-1}  \sigma^2_0(X)].
$

The following simple example serves to illustrate the robustness property in Corollary~2.
 For brevity we consider estimation of the treatment component $\beta_1 = E[Y(1)]$.
 \begin{example}[Model flexibility] \label{ex:1}
 Let $X_1$ and $X_2$ be independent variables from a beta distribution with both shape parameters equal $1/2$, 
 and define outcome variables as $Y(1) = \beta_1(X) + \epsilon_1$, where the error $\epsilon_1 \sim N(0,1/2)$ and the conditional mean
 \[
 \beta_1(X)  =  (X_1+1)^2 +\frac{1}{X_1+0.2}   + \log(X_2 + 0.7)   + 2e^{X_2}  - I(X_1>0.3) + I(X_2>0.6). 
 \]
 Three different treatment variables $T_1, T_2$, and $T_3$ are examined, which have binomial distributions conditional on $X$ specified by the propensity scores
 \[
 \begin{array}{l}
 e_{T_1}(X)  = e^{-1.2 - (X_1-X_2)},  \\
 e_{T_2}(X)  =  (X_2-X_1)^3/5 + (X_2-X_1)/5 + 1/2,  \\
 e_{T_3}(X)  = 1/( 3X_1-X_2 + 2.2).
 \end{array}
 \]

We define estimators $\esto{\kl,1}$ and $\esto{\qd,1}$ of $\beta_1$ by balancing $X_1$ and $X_2$, i.e., we use $U_2(X) = (X_1,X_2)$.  
 Even though the conditional mean outcome of the treated $\beta_1(X)$ is nonlinear in $X_1$ and $X_2$, it holds that  $\hat{\beta}_{\mbox{\scriptsize \kl},1}$ is consistent when the treatment variable is $T_1$, 
 and $\esto{\qd,1}$ is consistent when the treatment variable is $T_3$, as implied by the functional forms of the propensity scores $e_{T_1}(X)$ and $e_{T_3}(X)$, see Corollary 2. To confirm this, we compute approximations of the asymptotic errors $\tilde{\Delta}_{\mbox{\scriptsize \kl}}$ and $\tilde{\Delta}_{\mbox{\scriptsize \qd}}$ using the covariance representation from Corollary 1. First the limiting parameters $\tilde{\lambda}_1$ and $\tilde{\kappa}_1$ are estimated using the dual parameters $\hat{\lambda}_1$ and $\hat{\kappa}_1$ obtained from two large samples, and then the covariances $\tilde{\Delta}_{\mbox{\scriptsize \kl}}=Cov(Z^{\mbox{\scriptsize \kl}}_1(X),\beta_1(X))$ and $\tilde{\Delta}_{\mbox{\scriptsize \qd}}=Cov(Z^{\mbox{\scriptsize \qd}}_1(X),\beta_1(X))$ are approximated using sample covariances based on new independent sample, with the estimates of the dual limits plugged in, see Figure \ref{fig:Ex1}.
 \begin{figure}
  \centerline{\includegraphics[width=\textwidth]{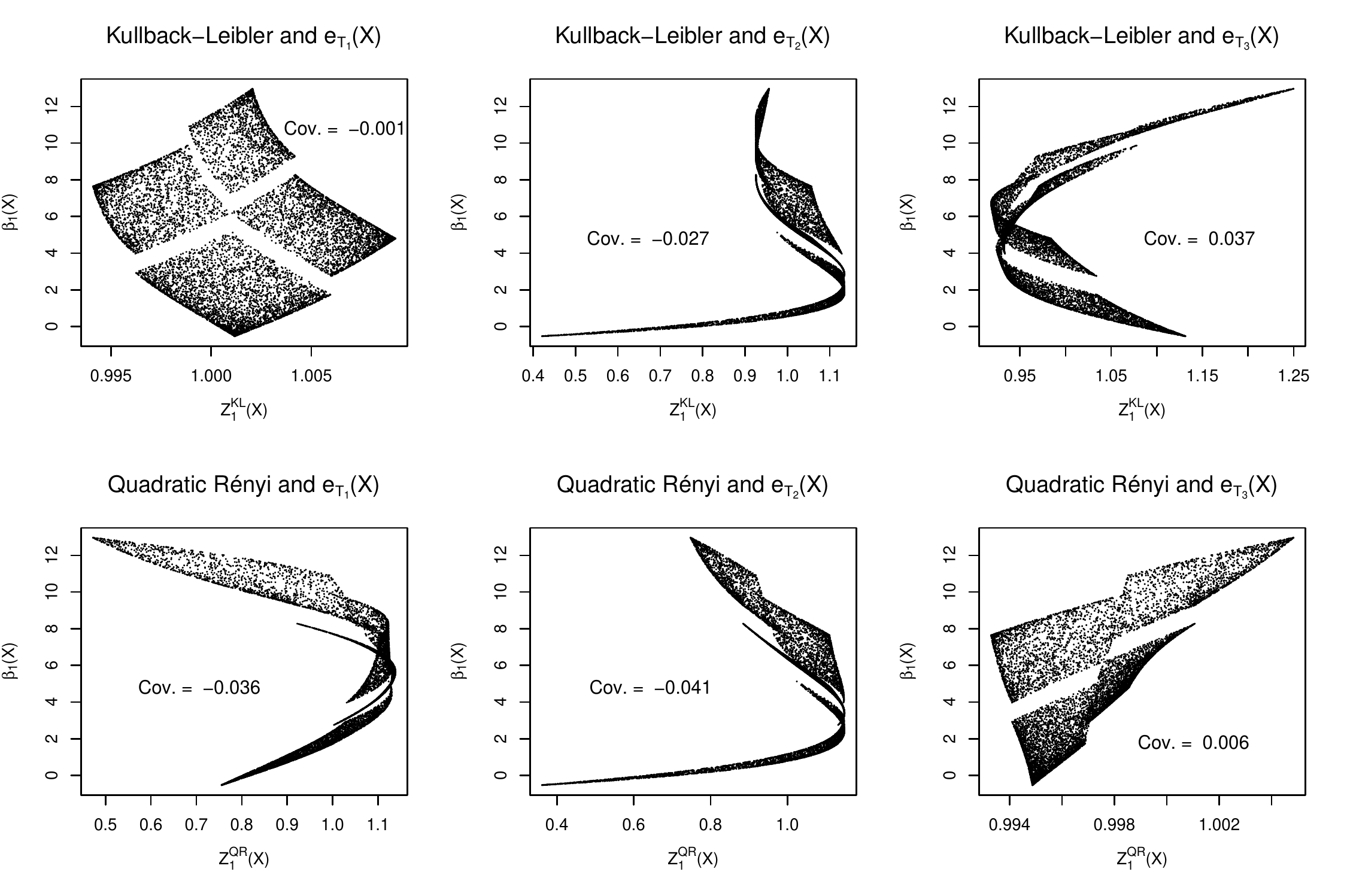}}
 \caption{
 Samples of size $N=10,000$ of the random vectors $(\beta_1(X), Z^{\mbox{\scriptsize \kl}}_1(X))$ and $(\beta_1(X), Z^{\mbox{\scriptsize \qd}}_1(X))$ for the  propensity scores $e_{T_j}(X), j = 1,2,3,$ as described in Example \ref{ex:1}. 
 }
 \label{fig:Ex1}
 \end{figure}
 \end{example}

\subsection{Variance estimation}  \label{sec:variance}
To make use of the normal approximation in Theorem \ref{th:norm}, we need estimators for the asymptotic variances $\tau^2_{\mbox{\scriptsize \kl}}$ and $\tau^2_{\mbox{\scriptsize \qd}}$.
We use the plug-in principle to construct estimators $\hat{\tau}^2_{{\mbox{\scriptsize \kl}}}$ and $\hat{\tau}^2_{{\mbox{\scriptsize \qd}}}$ which, under the assumption of constant conditional variances, $\sigma^2_t(X) = \sigma_t^2$, have the form
\begin{equation}\label{var:est}
\hat{\tau}^2_{{\mbox{\scriptsize \kl}}} = \widehat{Var}(\beta_1(X)-\beta_0(X)) + 
\hat{\sigma}^2_1 \frac{1}{N} \sum_{\{i: T_i = 1\}} (w^{\mbox{\scriptsize \kl}}_{1,i})^2 + 
\hat{\sigma}^2_0 \frac{1}{N} \sum_{\{i: T_i = 0\}} (w^{\mbox{\scriptsize \kl}}_{0,i})^2,
\end{equation}
and similar for $\hat{\tau}^2_{{\mbox{\scriptsize \qd}}}$.
Here, different nonparametric estimators of the regression functions can be used for estimation of the variance of the difference of conditional expectations, the first term in \eqref{var:est}, and the conditional variances $\hat{\sigma}^2_1,\hat{\sigma}^2_0$, see e.g., \citeN{ruppert1997local} and \citeN{fan1998efficient}. In the simulation study we use a simple approach with linear least squares regression models. In this case, the variance estimators are consistent under the assumptions of Theorem \ref{th:norm} (and $\sigma^2_t(X) = \sigma_t^2$). To see this, note that the two last terms in \eqref{var:est} converge weakly to the corresponding expected values, which can be verified by the limiting result for the dual parameters in Proposition~\ref{prop:1} together with a weak law of large numbers for averages with estimated parameters \cite[Th.\ 7.3]{SB:13}.  

\section{Simulations}\label{sec5}
The R code used for the simulations are available online, see \citeN{github}.
\subsection{Design}
We study an extension of the simulation scenario used in \citeN{hainmueller2012entropy}, which in turn originates from \citeN{frolich2007propensity}. 
This includes a six-dimensional covariate vector, $X = (X_1,X_2,X_3,X_4,X_5,X_6)$, where
$(X_1,X_2,X_3)$ is multivariate normal with mean zero and $Var(X_1) = 2, Var(X_2) = Var(X_3) = 1, Cov(X_1,X_2) = 1, Cov(X_1,X_3) = -1, Cov(X_2,X_3) = -1/2$, and where $X_4 = Unif(-3,3)$, $X_5 = \chi^2(df=1)$, $X_6 = Ber(0.5)$.

The treatment variable is defined as
$
T = \mathbf I (X_1 + 2X_2 - 2 X_3 - X_4 - 0.5 X_5 + X_6  + \epsilon > 0), 
$
and where the error term obeys $\epsilon \sim N(0, 30)$.
The outcome variables follow conditional normal models; $Y(t) = \beta_t(X) + \epsilon_t$, where $\epsilon_1$ and $\epsilon_0$ are independent and $\epsilon_t \sim N(0,1)$. 
Since the design in \citeN{hainmueller2012entropy} considers only the true treatment effect fixed at zero for all units we additionally use three different designs for the conditional mean $\beta_t(X)$, see Table \ref{tab:designs} for details.
\subsection{Estimators}
Both the Kullback-Leibler and the quadratic R\'{e}nyi divergences are examined, along with two variations of the balancing vector $U_K(X)$, given by $U_6(X)  =  (X_1,X_2,X_3,X_4,X_5,X_6)$, and a vector $U_{26}(X)$ that contains  $U_6(X)$ and all second order terms and interactions. Hence by the combinations of divergence measures and balancing vectors, four variations of entropy balancing estimators are included. We use the notation $\hat{\beta}^{(r)}, r=1,2$, for reference to which balancing vector that is used, either $U_6(X)$ ($r=1$) or $U_{26}(X)$ ($r=2$). Approximations of the asymptotic error $\tilde{\Delta}_{\mbox{\scriptsize \eb}}$ in Corollary \ref{cor:1} for the entropy balancing estimators in the considered simulation designs are also presented. The values of $\tilde{\Delta}_{\mbox{\scriptsize \eb}}$ are obtained as described in Example \ref{ex:1} (here we use $10^6$ samples and average over 10 replications).

For comparison, we also include three IPW-estimators of type \eqref{ipw}, based on the probit specifications 
\begin{itemize}
\item[(1)] $\alpha_0 + \alpha_1 X_1 + \alpha_2 X_2 + \alpha_3 X_3 + \alpha_4 X_4 + \alpha_5 X_5 + \alpha_6 X_6 $ (correct PS-model) 
\item[(2)]  $\alpha_0 + \alpha_1 X_1^2 + \alpha_2 X_2^2 + \alpha_3 X_3 + \alpha_4 X_4^2 + \alpha_5 X_5^2 + \alpha_6 X_6 $, 
\item[(3)]  $\alpha_0 + \alpha_1 X_1 X_3 + \alpha_2 X_2^2 +  \alpha_4 X_4 + \alpha_5 X_5 + \alpha_6 X_6$.
\end{itemize}

\subsection{Variance estimators} 
Three different methods for estimating the variances of and $\esto{\kl}$ and $\esto{\qd}$ are investigated. In addition to the proposed parametric OLS estimator (\ref{var:est}), we consider two nonparametric methods; an estimate of the efficiency bound (described below) and a simple bootstrap approach based on 10,000 resamples. 

As discussed in previous section, $\esto{\kl}$ and $\esto{\qd}$ can be regarded as nonparametric estimators if the balancing vector is assumed to grow in dimension with $N$, and then, under certain assumptions, the variances of $\esto{\kl}$ and $\esto{\qd}$ converge to the efficiency bound $\sigma^2_{eff}$. \citeN{chan2016globally} introduce a non-parametric estimator of $\sigma^2_{eff}$ which is an empirical counterpart of the representation $\sigma^2_{eff} = E[\psi(X,T,Y)^2],$
where 
$\psi(X,T,Y)$ is the {\it efficient influence function} given by
$$\psi(X,T,Y) = \frac{T Y}{e(X)} - \frac{(1-T) Y}{1-e(X)} - \beta - \beta_1(X) \left(\frac{T}{e(X)} - 1 \right) + 
\beta_0(X)\left(\frac{1-T}{1- e(X)} -1 \right).
$$
The estimator can be written as $ \hat{\sigma}^2_{eff} = N^{-1} \sum_{i=1}^N \hat{\psi}(X_i,T_i,Y_i)^2,$
where $\hat{\psi}(X,T,Y)$ is obtained by inserting estimates of the conditional mean $\beta_t(X),t=1,0$, and the inverse probabilities $e(X)^{-1}$ and $(1-e(X))^{-1}$. 
The estimate of $\beta_t(X)$, denoted by $\hat{\beta}_{t,w}(X)$, is
the weighted least squares estimator that regresses the outcome $Y$ on $U_K(X)$ in group $t=1,0$. For estimation of the inverse probabilities $\{ e(X_i)^{-1}, T_i = 1 \}$ and $\{(1 - e(X_i))^{-1}, T_i = 0\}$, the corresponding sets of weights $\{ w_{1,i}, T_i = 1 \}$ and $\{ w_{0,i}, T_i = 0 \}$ are used. We refer to the paper for more details on this estimator.

To indicate which variance estimator is used, we let P denote the proposed parametric OLS estimator (\ref{var:est}), NP denote the nonparametric estimator $\hat{\sigma}^2_{eff}$, and B denote the bootstrap estimator. For evaluation we use the ratio $R_M= \left (M^{-1}\sum_{i=1}^{M} \hat{\sigma}^2_i \right) / s^2_{M}$, where
 $s^2_{M}=(M-1)^{-1}\sum_{j=1}^{M}(\hat{\beta}_j-\overline{\hat{\beta}})^2$ is the Monte Carlo variance. 
 The ratio is interpreted as the proportion of the mean of estimated variances in relation to the true variance and we use a large sample approximation of its standard error \cite[Eq.\ 9.4]{SB:13}. 

\subsection{Results}
The bias, standard deviation and root mean squared error (RMSE) of the estimators are displayed in Table \ref{restab1} ($N=500$) and Table \ref{restab4} ($N=1000$). In addition, Table~\ref{restab2} ($N=500$) and Table~\ref{restab5} ($N=1000$) provide coverage, empirical standard errors, and means of estimated standard errors ($\hat{\tau}$) for three different variance estimators. Evaluations of the variance estimators are presented in Table \ref{restab3} ($N=500$) and Table \ref{restab6} ($N=1000$). The error approximations displayed in Table \ref{asymtab} are consistent with the observed bias for the EB estimators.

 The small bias of $\est{(1)}{\kl}$, $\est{(1)}{\qd}$, $\est{(2)}{\kl}$ and $\est{(2)}{\qd}$ in Design A, and of $\est{(2)}{\kl}$ and $\est{(2)}{\qd}$ in Designs B and C, is expected as the assumption of Corollary \ref{th:rob}$(i)$ is met for both balancing vectors $U_6(X)$ and $U_{26}(X)$ in Design A, and for $U_{26}(X)$ in Designs B and C (approximately for Design B). We get a small bias for the correctly specified IPW1, whereas the two misspecified IPW estimators are heavily biased. All three IPW estimators have relatively large standard error. In Design C the standard errors of $\est{(1)}{\kl}$ and $\est{(1)}{\qd}$ increase with factors around 8 and 3 compared to $\est{(2)}{\kl}$ and $\est{(2)}{\qd}$.

The results for the parametric variance estimators show that the means of the estimated variances are relatively close to the empirical for all four EB estimators. The bootstrap estimator has a similar performance, although it has a tendency to underestimate the variance of $\est{(1)}{\kl}$ in Design C, and to overestimate the variances of $\est{(2)}{\kl}$ and $\est{(2)}{\qd}$ for the smaller sample size $N=500$. The nonparametric variance estimator underestimates the variance in several cases, and this is more pronounced for the estimator  $\est{(2)}{\kl}$.
 
 Next we give separate comments for the two outcome designs no treatment effect and heterogeneous treatment effect
 \subsubsection{No treatment effect}
The bias is small for $\est{(1)}{\kl}$ and $\est{(1)}{\qd}$ in Design B and C, which is unexpected given that the assumption of Corollary \ref{th:rob}(i) is not satisfied for the shorter balancing vector $U_6(X)$. In Appendix B we provide a thorough examination of this result.
 
The interval coverage is close to $95\%$ for all estimators, with the exception of $\est{(2)}{\kl}$ when the nonparametric variance estimator is used. This is caused by underestimation of the variance. 
The same pattern is seen for both sample sizes, although for $N=500$ the coverage varies slightly more.

\subsubsection{Heterogeneous treatment effect}
We see that $\est{(1)}{\qd}$ and $\est{(1)}{\kl}$ are biased in Design B, and heavily biased in Design C. Note also that $\est{(1)}{\qd}$ is clearly worse, see the discussion in Appendix B. The coverage of $\est{(1)}{\kl}$ and $\est{(1)}{\qd}$ is lower than 95$\%$ in Designs B and C for all three variance estimators, where $\est{(1)}{\qd}$ is clearly worse. This is expected given their bias. Note that in Design~C, the variance of $\est{(1)}{\qd}$ is underestimated by the nonparametric- and bootstrap methods, but not by the parametric approach, which is reflected in the differences in coverage. For $\est{(2)}{\kl}$ and $\est{(2)}{\qd}$, the coverage is close to 95$\%$ in most cases although $\est{(2)}{\kl}$ combined with the nonparametric variance estimator has a lower coverage in Design B, due to underestimation of the variance.

\begin{table}
\caption{Simulation designs}
\begin{center}
\begin{tabular}{ l l }
OR-scenario & Coefficients\\
\hline
{\it Design A}   &   \\
Regressors: & \\
$R(X) = (1,X_1,X_2,X_3,X_4,X_5,X_6)$&\\
1) No treatment effect &  $\alpha = (0,1,1,1,-1,1,1)$ \\ 
{\small $\beta_1(X) = \beta_0(X) = \alpha^T R(X)$ } & \\ 
(True $\beta$ = 0, unadjusted $\beta$ $\approx 2.4 $) & \\ [1em]
2) Heterogenous treatment effect &  $\alpha_1 = (0.5,1.7,1.5,1.4,-0.5, 1.3,2.1)$ \\
{\small $\beta_1(X) = \alpha_1^T R(X)$, $\beta_0(X) = \alpha_0^T R(X)$}  & $\alpha_0 = (-0.7,0.7,0.3,0.8,-1.6,0.9,0.5)$ \\ 
(True $\beta = 2.4$, unadjusted $\beta$ $\approx 4.3 $) & \\ [1em]
\hline
{\it Design B} &    \\
Regressors: & \\
 $R(X) = (1,X_1,X_2,X_3 X_4,\sqrt{X_5})$ &\\
1) No treatment effect &  $\alpha = (0,1,1,0.2,-1)$ \\
(True $\beta$ = 0, unadjusted $\beta$ $\approx 2.2 $) & \\ [1em]
2) Heterogenous treatment effect &  $\alpha_1 = (-0.7, 1.7, 0.7, -0.2, -1.5)$ \\
 (True $\beta = -2$ , unadjusted $\beta$ $\approx 0 $) & $\alpha_0 = (0.6, 0.6, 0.5, 0.3, -0.5)$ \\ 
\hline
{\it Design C} &\\
Regressors: & \\
 $R(X) = (1,X_1^2,X_2^2, X_5^2,X_1 X_2, X_1 X_5, X_2 X_5)$ & \\
1) No treatment effect &  $\alpha = (0, 1,1,1,2,2,2)$ \\ 
 (True $\beta$ = 0, unadjusted $\beta$ $\approx 3.0 $) \\ [1em]
2) Heterogenous treatment effect &  $\alpha_1 = (0.6, 1.4, 1.5, 1.3, 2.5, 2.7, 2.2)$ \\
 (True $\beta = 6.5$, unadjusted $\beta$ $\approx 9.6 $) &  $\alpha_0 = (-0.4,  0.7, 0.5, 0.6, 1.5, 1.9, 1.6)$ \\ [1em]
\hline
\end{tabular}
\end{center}
\label{tab:designs}
\end{table}

\begin{table}
\caption{Asymptotic error $\tilde{\Delta}_{\mbox{\scriptsize \eb}}$ in Corollary \ref{cor:1} for the simulation designs. Approximate values; mean over 10 simulations and corresponding standard errors.}
\begin{tabular}{crccc}\label{asymtab}
& & Design A & Design B & Design C \\  
 \parbox[t]{4mm}{\multirow{4}{*}{\rotatebox[origin=c]{90}{\shortstack{\footnotesize No treatment \\ \footnotesize effect}}}}  
  &$\est{(1)}{\kl}$ & 0.001(0.002) & -0.001(0.002) & 0.011(0.01) \\ 
  &$\est{(1)}{\qd}$ & 0.001(0.002) & -0.001(0.001) & -0.005(0.011) \\ 
  &$\est{(2)}{\kl}$ & -0.004(0.002) & -0.006(0.004) & 0.031(0.018) \\ 
  &$\est{(2)}{\qd}$ & 0.001(0.003) & 0.001(0.003) & 0.016(0.017) \\ [1em]
  
 \parbox[t]{4mm}{\multirow{4}{*}{\rotatebox[origin=c]{90}{\shortstack{\footnotesize Heterogeneous \\ \footnotesize treatment effect}}}}  
  &$\est{(1)}{\kl}$ & 0.000(0.002) & 0.081(0.002) & -0.788(0.009) \\ 
  &$\est{(1)}{\qd}$ & -0.001(0.002) & 0.151(0.001) & -1.491(0.012) \\ 
  &$\est{(2)}{\kl}$ & 0.005(0.005) & -0.003(0.003) & -0.014(0.019) \\ 
  &$\est{(2)}{\qd}$ & 0.001(0.003) & 0.002(0.002) & -0.022(0.011) 
\end{tabular}
\end{table}

\begin{table}
\caption{
\caps{500}
}
\centering
\begin{tabular}{ll ccc|ccc|ccc}\label{restab1}
&& \multicolumn{3}{c}{Design A}  & \multicolumn{3}{c}{Design B} & \multicolumn{3}{c}{Design C} \\
 && BIAS & SD & RMSE & BIAS & SD & RMSE & BIAS & SD & RMSE \\ 
     \parbox[t]{2mm}{\multirow{7}{*}{\rotatebox[origin=c]{90}{No treatment effect}}}
 & $\est{(1)}{\kl}$   & -0.003 & 0.111 & 0.111 & 0.002 & 0.122 & 0.122 & 0.060 & 1.140 & 1.141 \\ 
 & $\est{(1)}{\qd}$   & -0.004 & 0.108 & 0.108 & 0.002 & 0.116 & 0.116 & 0.027 & 1.118 & 1.118 \\ 
 & $\est{(2)}{\kl}$   & 0.000 & 0.139 & 0.139 & 0.010 & 0.148 & 0.148 & 0.004 & 0.147 & 0.147 \\ 
 & $\est{(2)}{\qd}$   & -0.002 & 0.131 & 0.131 & 0.010 & 0.138 & 0.138 & 0.003 & 0.137 & 0.137 \\ 
 & IPW1 			  & 0.049 & 0.350 & 0.353 & 0.065 & 0.396 & 0.401 & 0.020 & 2.380 & 2.379 \\ 
 & IPW2 			  & 1.838 & 0.282 & 1.859 & 1.131 & 0.226 & 1.153 & 2.164 & 1.013 & 2.389 \\ 
 & IPW3				  & 1.335 & 0.177 & 1.347 & 2.227 & 0.208 & 2.237 & 4.468 & 1.288 & 4.650 \\ 
 [1em]
 \parbox[t]{7mm}{\multirow{7}{*}{\rotatebox[origin=c]{90}{\shortstack{Heterogeneous \\ treatment effect}}}}& 
 $\est{(1)}{\kl}$      & 0.005 & 0.176 & 0.176 & 0.076 & 0.147 & 0.165 & -0.822 & 1.255 & 1.500 \\ 
 & $\est{(1)}{\qd}$    & 0.005 & 0.174 & 0.174 & 0.144 & 0.147 & 0.206 & -1.508 & 1.219 & 1.939 \\ 
 & $\est{(2)}{\kl}$    & 0.004 & 0.195 & 0.195 & 0.010 & 0.169 & 0.169 & -0.021 & 0.421 & 0.422 \\ 
 & $\est{(2)}{\qd}$    & 0.005 & 0.190 & 0.190 & 0.006 & 0.160 & 0.160 & -0.021 & 0.416 & 0.416 \\ 
 & IPW1 			   & 0.046 & 0.424 & 0.426 & 0.076 & 0.425 & 0.431 & -0.058 & 2.485 & 2.484 \\ 
 & IPW2 			   & 1.941 & 0.339 & 1.970 & 1.079 & 0.237 & 1.105 & 1.759 & 1.094 & 2.071 \\ 
 & IPW3 			   & 1.428 & 0.225 & 1.446 & 2.142 & 0.208 & 2.152 & 4.845 & 1.433 & 5.052 
\end{tabular}
\end{table}

\begin{table}
\caption{\capS{500}}
\centering
\begin{tabular}{ll ccc|ccc|ccc}\label{restab2}
&  & \multicolumn{3}{c}{Design A}  & \multicolumn{3}{c}{Design B}  & \multicolumn{3}{c}{Design C} \\
& & Cov. & $\sqrt{N}\mbox{SD}$ & $\overline{\hat{\tau}}$ & Cov. & $\sqrt{N}\mbox{SD}$ & $\overline{\hat{\tau}}$ & Cov. & $\sqrt{N}\mbox{SD}$ & $\overline{\hat{\tau}}$ \\ 
  \parbox[b]{2mm}{\multirow{12}{*}{\rotatebox[origin=b]{90}{ No treatment effect}}} 
&   $\est{(1)}{\kl}$-P    & 0.95 & 3.51 & 3.54 & 0.94 & 3.87 & 3.81 & 0.95 & 36.06 & 36.35 \\ 
&   $\est{(1)}{\qd}$-P           & 0.96 & 3.41 & 3.41 & 0.95 & 3.68 & 3.68 & 0.94 & 35.36 & 35.27 \\ 
&	$\est{(2)}{\kl}$-P           & 0.96 & 4.39 & 4.51 & 0.94 & 4.68 & 4.55 & 0.95 & 4.66 & 4.55 \\ 
&	$\est{(2)}{\qd}$-P           & 0.96 & 4.15 & 4.22 & 0.94 & 4.35 & 4.27 & 0.95 & 4.33 & 4.25 \\ 
& $\est{(1)}{\kl}$-NP    & 0.93 & 3.51 & 3.39 & 0.92 & 3.87 & 3.62 & 0.91 & 36.06 & 30.78 \\ 
&   $\est{(1)}{\qd}$-NP           & 0.94 & 3.41 & 3.36 & 0.95 & 3.68 & 3.62 & 0.93 & 35.36 & 33.16 \\ 
&	$\est{(2)}{\kl}$-NP           & 0.87 & 4.39 & 3.33 & 0.84 & 4.68 & 3.38 & 0.84 & 4.66 & 3.36 \\ 
&	$\est{(2)}{\qd}$-NP           & 0.93 & 4.15 & 3.86 & 0.92 & 4.35 & 3.91 & 0.93 & 4.33 & 3.89 \\ 
& $\est{(1)}{\kl}$-B    & 0.94 & 3.51 & 3.47 & 0.93 & 3.87 & 3.72 & 0.92 & 36.06 & 32.19 \\ 
&   $\est{(1)}{\qd}$-B           & 0.95 & 3.41 & 3.40 & 0.95 & 3.68 & 3.66 & 0.93 & 35.36 & 33.63 \\ 
&	$\est{(2)}{\kl}$-B           & 0.96 & 4.39 & 4.73 & 0.95 & 4.68 & 4.84 & 0.96 & 4.66 & 4.89 \\ 
&	$\est{(2)}{\qd}$-B           & 0.96 & 4.15 & 4.41 & 0.95 & 4.35 & 4.48 & 0.96 & 4.33 & 4.44 \\ 
[1em]
\parbox[b]{7mm}{\multirow{12}{*}{\rotatebox[origin=c]{90}{\shortstack{ Heterogeneous treatment effect}}}} 
& $\est{(1)}{\kl}$   	 & 0.94 & 5.56 & 5.52 & 0.92 & 4.64 & 4.69 & 0.87 & 39.70 & 39.75 \\ 
&$\est{(1)}{\qd}$-P  				 & 0.94 & 5.51 & 5.43 & 0.82 & 4.65 & 4.57 & 0.76 & 38.55 & 38.74 \\ 
&$\est{(2)}{\kl}$-P  				 & 0.95 & 6.18 & 6.20 & 0.95 & 5.34 & 5.39 & 0.93 & 13.33 & 12.98 \\ 
&$\est{(2)}{\qd}$-P 				 & 0.94 & 6.00 & 5.98 & 0.95 & 5.06 & 5.13 & 0.93 & 13.16 & 12.87 \\ 
&  $\est{(1)}{\kl}$-NP   	 & 0.94 & 5.56 & 5.41 & 0.91 & 4.64 & 4.53 & 0.80 & 39.70 & 34.64 \\ 
&$\est{(1)}{\qd}$-NP  				 & 0.94 & 5.51 & 5.39 & 0.82 & 4.65 & 4.57 & 0.71 & 38.55 & 36.62 \\ 
&$\est{(2)}{\kl}$-NP  				 & 0.93 & 6.18 & 5.39 & 0.90 & 5.34 & 4.42 & 0.92 & 13.33 & 12.54 \\ 
&$\est{(2)}{\qd}$-NP 				 & 0.93 & 6.00 & 5.73 & 0.94 & 5.06 & 4.84 & 0.93 & 13.16 & 12.74 \\ 
& $\est{(1)}{\kl}$-B   	 & 0.94 & 5.56 & 5.47 & 0.91 & 4.64 & 4.62 & 0.81 & 39.70 & 36.09 \\ 
&$\est{(1)}{\qd}$-B  				 & 0.94 & 5.51 & 5.42 & 0.83 & 4.65 & 4.61 & 0.71 & 38.55 & 37.01 \\ 
&$\est{(2)}{\kl}$-B  				 & 0.95 & 6.18 & 6.39 & 0.96 & 5.34 & 5.69 & 0.93 & 13.33 & 13.09 \\ 
&$\est{(2)}{\qd}$-B 				 & 0.95 & 6.00 & 6.12 & 0.95 & 5.06 & 5.32 & 0.93 & 13.16 & 12.93 \\ 
\end{tabular}
\end{table}

\begin{table}
\caption{Variance estimators; parametric (P), nonparametric (NP), bootstrap (B), sample size $N=500$, and $M= 1000$ replications. Ratio $R_M = \left (M^{-1}\sum_{i=1}^{M} \hat{\sigma}^2_i \right) / s^2_{M}$, with Monte Carlo variance  $s^2_{M}$, and standard error of $R_M$.
}
\centering
\begin{tabular}{llcccccc}\label{restab3}
 && \multicolumn{2}{c}{Design A}&\multicolumn{2}{c}{Design B} &\multicolumn{2}{c}{Design C} \\
 && $R_M$ & $sd(R_M)$ & $R_M$ & $sd(R_M)$ & $R_M$ & $sd(R_M)$ \\ 
 \parbox[b]{4mm}{\multirow{12}{*}{\rotatebox[origin=b]{90}{ No treatment effect}}} 
 & $\est{(1)}{\kl}$-P         & 1.02 & 0.05 & 0.97 & 0.05 & 1.04 & 0.05 \\ \ 
 & $\est{(1)}{\qd}$-P          & 1.00 & 0.05 & 1.00 & 0.05 & 1.02 & 0.05 \\ 
 & $\est{(2)}{\kl}$-P          & 1.07 & 0.05 & 0.96 & 0.04 & 0.96 & 0.04 \\ 
 & $\est{(2)}{\qd}$-P          & 1.04 & 0.05 & 0.97 & 0.04 & 0.97 & 0.04 \\ 
 & $\est{(1)}{\kl}$-NP       & 0.94 & 0.04 & 0.88 & 0.04 & 0.76 & 0.04 \\ 
 & $\est{(1)}{\qd}$-NP       & 0.97 & 0.05 & 0.97 & 0.04 & 0.91 & 0.04 \\ 
 & $\est{(2)}{\kl}$-NP       & 0.58 & 0.03 & 0.53 & 0.02 & 0.53 & 0.02 \\ 
 & $\est{(2)}{\qd}$-NP       & 0.87 & 0.04 & 0.81 & 0.04 & 0.82 & 0.04 \\ 
 & $\est{(1)}{\kl}$-B     & 0.98 & 0.05 & 0.93 & 0.04 & 0.84 & 0.04 \\ 
 & $\est{(1)}{\qd}$-B     & 0.99 & 0.05 & 0.99 & 0.05 & 0.94 & 0.04 \\ 
 & $\est{(2)}{\kl}$-B     & 1.18 & 0.06 & 1.10 & 0.05 & 1.14 & 0.05 \\ 
 & $\est{(2)}{\qd}$-B     & 1.14 & 0.05 & 1.07 & 0.05 & 1.06 & 0.05 \\ 
 [1em]
 \parbox[t]{4mm}{\multirow{12}{*}{\rotatebox[origin=c]{90}{ Heterogeneous treatment effect}}}   
  &  $\est{(1)}{\kl}$-P   & 0.98 & 0.04 & 1.02 & 0.04 & 1.03 & 0.05 \\ 
  & $\est{(1)}{\qd}$-P    & 0.97 & 0.04 & 0.97 & 0.04 & 1.03 & 0.05 \\ 
  & $\est{(2)}{\kl}$-P    & 1.01 & 0.05 & 1.03 & 0.05 & 0.98 & 0.04 \\ 
  & $\est{(2)}{\qd}$-P    & 1.00 & 0.05 & 1.03 & 0.05 & 0.99 & 0.04 \\ 
  & $\est{(1)}{\kl}$-NP   & 0.95 & 0.04 & 0.96 & 0.04 & 0.80 & 0.04 \\ 
  & $\est{(1)}{\qd}$-NP   & 0.96 & 0.04 & 0.97 & 0.04 & 0.93 & 0.04 \\ 
  & $\est{(2)}{\kl}$-NP   & 0.77 & 0.04 & 0.70 & 0.03 & 0.92 & 0.04 \\ 
  & $\est{(2)}{\qd}$-NP   & 0.92 & 0.04 & 0.92 & 0.04 & 0.98 & 0.04 \\ 
  & $\est{(1)}{\kl}$-B    & 0.97 & 0.04 & 1.00 & 0.04 & 0.87 & 0.05 \\ 
  & $\est{(1)}{\qd}$-B    & 0.97 & 0.04 & 0.99 & 0.04 & 0.95 & 0.05 \\ 
  & $\est{(2)}{\kl}$-B    & 1.08 & 0.05 & 1.16 & 0.05 & 1.00 & 0.04 \\ 
  & $\est{(2)}{\qd}$-B    & 1.05 & 0.05 & 1.11 & 0.05 & 1.00 & 0.04 
\end{tabular}
\end{table}

\begin{table}
\caption{
\caps{1000}
}
\centering
\begin{tabular}{cc ccc|ccc|ccc}\label{restab4}
&& \multicolumn{3}{c}{Design A}  & \multicolumn{3}{c}{Design B} & \multicolumn{3}{c}{Design C} \\
 && BIAS & SD & RMSE & BIAS & SD & RMSE & BIAS & SD & RMSE \\ 
     \parbox[t]{2mm}{\multirow{7}{*}{\rotatebox[origin=c]{90}{No treatment effect}}}
 & $\est{(1)}{\kl}$   & 0.004 & 0.078 & 0.078 & -0.003 & 0.085 & 0.085 & 0.024 & 0.779 & 0.779 \\ 
 & $\est{(1)}{\qd}$   & 0.004 & 0.074 & 0.074 & -0.004 & 0.083 & 0.083 & 0.005 & 0.765 & 0.764 \\ 
 & $\est{(2)}{\kl}$   & 0.003 & 0.095 & 0.095 & 0.003 & 0.096 & 0.097 & -0.001 & 0.096 & 0.096 \\ 
 & $\est{(2)}{\qd}$   & 0.004 & 0.089 & 0.089 & 0.001 & 0.089 & 0.089 & 0.000 & 0.091 & 0.091 \\ 
 & IPW1 						                  & 0.017 & 0.307 & 0.307 & 0.034 & 0.290 & 0.292 & 0.042 & 1.975 & 1.975 \\ 
 & IPW2 						                  & 1.854 & 0.204 & 1.865 & 1.124 & 0.157 & 1.135 & 2.256 & 1.005 & 2.469 \\ 
 & IPW3						                      & 1.341 & 0.123 & 1.347 & 2.231 & 0.141 & 2.235 & 4.460 & 0.783 & 4.528 \\ 
  [1em]
 \parbox[t]{7mm}{\multirow{7}{*}{\rotatebox[origin=c]{90}{\shortstack{Heterogeneous \\ treatment effect}}}}& 
 $\est{(1)}{\kl}$      & -0.001 & 0.123 & 0.123 & 0.079 & 0.106 & 0.133 & -0.792 & 0.928 & 1.220 \\ 
 & $\est{(1)}{\qd}$    & -0.000 & 0.120 & 0.120 & 0.148 & 0.105 & 0.181 & -1.488 & 0.889 & 1.733 \\ 
 & $\est{(2)}{\kl}$    & -0.001 & 0.133 & 0.133 & 0.006 & 0.118 & 0.118 & 0.013 & 0.297 & 0.297 \\ 
 & $\est{(2)}{\qd}$    & 0.000 & 0.130 & 0.130 & 0.005 & 0.113 & 0.113 & 0.012 & 0.295 & 0.295 \\ 
 & IPW1 						                   & 0.043 & 0.343 & 0.346 & 0.035 & 0.397 & 0.398 & 0.030 & 2.073 & 2.073 \\ 
 & IPW2 						                   & 1.960 & 0.238 & 1.974 & 1.077 & 0.177 & 1.092 & 2.058 & 1.332 & 2.451 \\ 
 & IPW3 						                   & 1.433 & 0.157 & 1.442 & 2.144 & 0.149 & 2.149 & 4.984 & 1.143 & 5.114 \\ 
\end{tabular}
\end{table}

\begin{table}
\caption{\capS{1000}}
\centering
\begin{tabular}{rl ccc|ccc|ccc}\label{restab5}
&  & \multicolumn{3}{c}{Design A}  & \multicolumn{3}{c}{Design B}  & \multicolumn{3}{c}{Design C} \\
& & Cov.\ & $\sqrt{N}\mbox{SD}$ & $\overline{\hat{\tau}}$ & Cov.\ & $\sqrt{N}\mbox{SD}$ & $\overline{\hat{\tau}}$ & Cov.\ & $\sqrt{N}\mbox{SD}$ & $\overline{\hat{\tau}}$ \\ 
  \parbox[b]{4mm}{\multirow{12}{*}{\rotatebox[origin=b]{90}{ No treatment effect}}} 
& $\est{(1)}{\kl}$-P    & 0.95 & 2.45 & 2.48 & 0.95 & 2.69 & 2.68 & 0.95 & 24.64 & 25.72 \\ 
&   $\est{(1)}{\qd}$-P           & 0.96 & 2.33 & 2.40 & 0.95 & 2.63 & 2.58 & 0.96 & 24.18 & 24.98 \\ 
&	$\est{(2)}{\kl}$-P           & 0.95 & 3.00 & 3.01 & 0.95 & 3.05 & 3.04 & 0.95 & 3.04 & 3.01 \\ 
&	$\est{(2)}{\qd}$-P           & 0.95 & 2.82 & 2.83 & 0.95 & 2.81 & 2.86 & 0.94 & 2.88 & 2.83 \\ 
& $\est{(1)}{\kl}$-NP & 0.95 & 2.45 & 2.43 & 0.94 & 2.69 & 2.61 & 0.94 & 24.64 & 23.08 \\  
&   $\est{(1)}{\qd}$-NP        & 0.95 & 2.33 & 2.38 & 0.95 & 2.63 & 2.57 & 0.95 & 24.18 & 23.90 \\  
&	$\est{(2)}{\kl}$-NP        & 0.90 & 3.00 & 2.49 & 0.89 & 3.05 & 2.52 & 0.89 & 3.04 & 2.50 \\ 
&	$\est{(2)}{\qd}$-NP        & 0.94 & 2.82 & 2.71 & 0.93 & 2.81 & 2.74 & 0.93 & 2.88 & 2.71 \\ 
&  $\est{(1)}{\kl}$-B  & 0.95 & 2.45 & 2.46 & 0.95 & 2.69 & 2.64 & 0.95 & 24.64 & 23.50 \\ 
&   $\est{(1)}{\qd}$-B         & 0.96 & 2.33 & 2.39 & 0.95 & 2.63 & 2.58 & 0.95 & 24.18 & 24.02 \\ 
&	$\est{(2)}{\kl}$-B         & 0.94 & 3.00 & 2.90 & 0.94 & 3.05 & 2.94 & 0.93 & 3.04 & 2.90 \\ 
&	$\est{(2)}{\qd}$-B         & 0.95 & 2.82 & 2.86 & 0.95 & 2.81 & 2.89 & 0.95 & 2.88 & 2.85 \\ 
[1em]  
\parbox[b]{4mm}{\multirow{12}{*}{\rotatebox[origin=c]{90}{ Heterogeneous treatment effect}}} 
& $\est{(1)}{\kl}$-P   	 & 0.95 & 3.88 & 3.88 & 0.88 & 3.36 & 3.29 & 0.84 & 29.33 & 28.56 \\ 
&$\est{(1)}{\qd}$-P  				 & 0.95 & 3.81 & 3.82 & 0.68 & 3.31 & 3.21 & 0.56 & 28.12 & 27.86 \\ 
&$\est{(2)}{\kl}$-P  				 & 0.95 & 4.20 & 4.24 & 0.95 & 3.73 & 3.63 & 0.94 & 9.40 & 9.27 \\ 
& $\est{(2)}{\qd}$-P 				 & 0.95 & 4.11 & 4.11 & 0.94 & 3.56 & 3.49 & 0.94 & 9.34 & 9.21 \\ 
& $\est{(1)}{\kl}$-NP  & 0.95 & 3.88 & 3.84 & 0.87 & 3.36 & 3.24 & 0.76 & 29.33 & 26.26 \\ 
&$\est{(1)}{\qd}$-NP  			& 0.95 & 3.81 & 3.81 & 0.69 & 3.31 & 3.23 & 0.52 & 28.12 & 26.71 \\ 
&$\est{(2)}{\kl}$-NP  			& 0.93 & 4.20 & 3.89 & 0.91 & 3.73 & 3.22 & 0.94 & 9.40 & 9.11 \\ 
&$\est{(2)}{\qd}$-NP 			& 0.95 & 4.11 & 4.03 & 0.94 & 3.56 & 3.39 & 0.94 & 9.34 & 9.17 \\ 
&  $\est{(1)}{\kl}$-B   & 0.95 & 3.88 & 3.86 & 0.88 & 3.36 & 3.27 & 0.77 & 29.33 & 26.73 \\ 
&$\est{(1)}{\qd}$-B  			& 0.95 & 3.81 & 3.82 & 0.69 & 3.31 & 3.25 & 0.52 & 28.12 & 26.81 \\ 
&$\est{(2)}{\kl}$-B  			& 0.95 & 4.20 & 4.17 & 0.94 & 3.73 & 3.57 & 0.94 & 9.40 & 9.24 \\ 
&$\est{(2)}{\qd}$-B 			& 0.96 & 4.11 & 4.13 & 0.94 & 3.56 & 3.52 & 0.94 & 9.34 & 9.21 
\end{tabular}
\end{table}

\begin{table}
\caption{Variance estimators; parametric (P), nonparametric (NP), bootstrap (B), sample size $N=1000$, and $M= 1000$ replications. Ratio $R_M = \left (M^{-1}\sum_{i=1}^{M} \hat{\sigma}^2_i \right) / s^2_{M}$, 
with Monte Carlo variance  $s^2_{M}$, and standard error of $R_M$.
}
\centering
\begin{tabular}{rlcccccc}\label{restab6}
 && \multicolumn{2}{c}{Design A}&\multicolumn{2}{c}{Design B} &\multicolumn{2}{c}{Design C} \\
 && $R_M$ & $sd(R_M)$ & $R_M$ & $sd(R_M)$ & $R_M$ & $sd(R_M)$ \\ 
 \parbox[b]{4mm}{\multirow{12}{*}{\rotatebox[origin=b]{90}{No treatment effect}}} 
 &  $\est{(1)}{\kl}$-P  & 1.03 & 0.05 & 0.99 & 0.05 & 1.10 & 0.05 \\ 
 & $\est{(1)}{\qd}$-P   & 1.06 & 0.05 & 0.96 & 0.04 & 1.08 & 0.05 \\ 
 & $\est{(2)}{\kl}$-P   & 1.01 & 0.04 & 1.00 & 0.05 & 0.99 & 0.04 \\ 
 & $\est{(2)}{\qd}$-P   & 1.01 & 0.05 & 1.04 & 0.05 & 0.97 & 0.04 \\ 
 & $\est{(1)}{\kl}$-NP  & 0.99 & 0.05 & 0.95 & 0.04 & 0.90 & 0.04 \\ 
 & $\est{(1)}{\qd}$-NP  & 1.04 & 0.05 & 0.95 & 0.04 & 1.00 & 0.04 \\ 
 & $\est{(2)}{\kl}$-NP  & 0.70 & 0.03 & 0.69 & 0.03 & 0.68 & 0.03 \\ 
 & $\est{(2)}{\qd}$-NP  & 0.93 & 0.04 & 0.95 & 0.04 & 0.88 & 0.04 \\ 
 & $\est{(1)}{\kl}$-B   & 1.01 & 0.05 & 0.97 & 0.04 & 0.94 & 0.04 \\ 
 & $\est{(1)}{\qd}$-B   & 1.05 & 0.05 & 0.96 & 0.04 & 1.00 & 0.05 \\ 
 & $\est{(2)}{\kl}$-B   & 0.95 & 0.04 & 0.94 & 0.04 & 0.92 & 0.04 \\ 
 & $\est{(2)}{\qd}$-B   & 1.03 & 0.05 & 1.06 & 0.05 & 0.98 & 0.04 \\ 
 [1em] 
 \parbox[b]{4mm}{\multirow{12}{*}{\rotatebox[origin=c]{90}{Heterogeneous treatment effect}}}   
  &  $\est{(1)}{\kl}$-P     & 1.00 & 0.04 & 0.96 & 0.04 & 0.96 & 0.06 \\ 
  & $\est{(1)}{\qd}$-P      & 1.01 & 0.04 & 0.94 & 0.04 & 1.00 & 0.05 \\ 
  & $\est{(2)}{\kl}$-P      & 1.02 & 0.04 & 0.95 & 0.04 & 0.99 & 0.05 \\ 
  & $\est{(2)}{\qd}$-P      & 1.00 & 0.04 & 0.96 & 0.04 & 0.99 & 0.05 \\ 
  & $\est{(1)}{\kl}$-NP   & 0.98 & 0.04 & 0.93 & 0.04 & 0.84 & 0.05 \\ 
  & $\est{(1)}{\qd}$-NP   & 1.00 & 0.04 & 0.95 & 0.04 & 0.93 & 0.05 \\ 
  & $\est{(2)}{\kl}$-NP   & 0.86 & 0.04 & 0.75 & 0.03 & 0.96 & 0.05 \\ 
  & $\est{(2)}{\qd}$-NP   & 0.96 & 0.04 & 0.91 & 0.04 & 0.99 & 0.05 \\ 
  & $\est{(1)}{\kl}$-B & 0.99 & 0.04 & 0.95 & 0.04 & 0.87 & 0.05 \\ 
  & $\est{(1)}{\qd}$-B & 1.01 & 0.04 & 0.96 & 0.04 & 0.94 & 0.05 \\ 
  & $\est{(2)}{\kl}$-B & 0.99 & 0.04 & 0.92 & 0.04 & 0.99 & 0.05 \\ 
  & $\est{(2)}{\qd}$-B & 1.01 & 0.04 & 0.98 & 0.04 & 1.00 & 0.05 
\end{tabular}
\end{table}

\begin{table}[ht]
\centering
\caption{
Estimates of the average effect $\beta$ in the diabetes data, 95$\%$ confidence intervals 
$\hat{\beta} \pm 1.96\,\hat{\sigma}(\hat{\beta})$, using a parametric (P) and nonparametric (NP) variance estimator, and an IPW estimator. 
}
\color{black}
\begin{tabular}{lcc}
  & $\hat{\beta}$ & 95$\%$ CI \\ 
   $\esto{\kl}$-P & 0.86 & (0.37, 1.35) \\ 
  $ \esto{\qd}$-P & 0.90 &  (0.42, 1.37)\\ 
  $\esto{\kl}$-NP &0.86 &(0.54, 1.17)\\
  $\esto{\qd}$-NP &0.90 &(0.51, 1.28)\\
   IPW,LOGIT & 0.81 & (0.49, 1.13) 
\end{tabular}
\end{table}

\section{Data example: Acute complications of Type 1 diabetes mellitus}\label{sec6}
The Swedish Childhood Diabetes Register is a population based register recording incident cases of Type 1 Diabetes Mellitus (T1DM) in children aged 0-14 years. T1DM is a chronic autoimmune disease which can lead to deadly complications both in short and long term. The key aspect for managing the disease is the regulation of blood sugar levels. Small children often rely on their parents but as they reach adolescence the main responsibility transfers to the youth themselves.
If the disease is poorly managed T1DM can cause acute complications such as, for example, ketoacidosis or coma. Our aim is to evaluate the effect of the individuals' ability to manage their disease on hospitalization due to accute complications of T1DM. Here, we define the ability by considering school achievements in the form of GPA of the individual, see similar studies by e.g., \citeN{almquist2013school,govan2012effect}. Addressing the manipulability of the treatment (low school grades), a possible intervention (target trial) could correspond to, for example, an intervention assigning extra tuition for a randomly chosen individual. In this observational study, we estimate the effect of having low GPA on being hospitalized with an acute complication of T1DM while controlling for confounding.
The study population consists of children, 0-14 year, diagnosed with T1DM between the years 1998 and 2006 (N=3761). Since socioeconomic factors are likely to be associated to both school achievements and health-related problems we control for them as confounders in the analysis. The access to treatment, outcome and confounders is facilitated by linkage from the SCDR to the Longitudinal Integration Database for Health Insurance and Labour Market Studies, the Inpatient Register, the Swedish Register of Education and the Multigenerational Register, making accessible a large pool of covariates for the children as well as their parents (259 individuals excluded from the study due to missingnes on some variables). We define the outcome as the number of times an individual has been hospitalized with the acute complications ketoacidosis, hyperosmolarity, lactic acidosis, hypoglycaemic
coma and/or coma, excluding hospitalizations where the secondary diagnosis is related to
pregnancy and other non-relevant diagnoses.  The number of hospitalizations, is measured after the grade is received with a 7-year follow-up. The treatment variable,  ”low grade”,  is defined as having received a grade lower than 165 (the 20th percentile). For more information on the dataset and variables, see Appendix C. A pool of 22 covariates $U_K(X) = (u_1(X),\ldots, u_{22}(X))$ are balanced; gender, age at onset of T1DM and socioeconomic characteristics of the parents. The socioeconomic variables are measured one year prior to receiving the grade, see the description provided in the supplementary material. The unadjusted mean outcome difference is 0.96 hospitalizations, with an average observed outcome of 1.54 (0.59) observed among those with low (high) grades. The results show that the estimated average causal effect of having low grades on days at the hospital due to acute complications of T1DM is $\esto{\qd}=0.90$  with 95\% confidence interval from 0.42 to 1.37, the estimator $\esto{\kl}$ gives an estimated causal effect of 0.86 days with 95\% confidence interval from 0.37 to 1.35. Using the variance estimator of \citeN{chan2016globally} instead gives the confidence intervals of (0.51, 1.28) for $\esto{\qd}$ and (0.54, 1.17) for $\esto{\kl}$. For comparison, a standard logistic regression IPW-estimator is applied yielding an estimate of $0.81$ with 95\% confidence interval of  (0.49, 1.13).

\section{Discussion}\label{sec7} 
The entropy balancing weights are obtained by an optimization procedure that directly targets two features: 
covariate balance to adjust for confounding,
and minimum divergence relative a set of uniform base weights to retain maximum amount of information.
Although the set of balancing functions and the divergence measure are chosen by the user the role of their specification for the estimator's properties has not been studied previously.
Here, we show that even though these components are not directly related to a treatment or outcome model, the conditions for consistency and asymptotic normality entail that implicit parametric model assumptions are made when we specify them. 
For consistency, we impose an outcome regression model for each group, or a combination of a propensity score model and an outcome regression for one of the groups. The balance functions form the regressors in the outcome- and propensity score models, and the link function in the propensity score model is determined by the divergence measure.

The entropy balancing estimators studied in this paper are not doubly robust in the conventional sense since a correctly specified propensity score model is not enough for consistency but needs to be combined with an outcome model for at least one of the groups. Although seemingly restrictive,  these are two alternative conditions, one for each group with different propensity score models, hereby offering some flexibility.

By construction, 
entropy balancing adjusts the sample in such a way that an empirical covariate balance condition is satisfied between the treated, control, and the combined group. This three-way balance is a distinctive feature of the method which can be contrasted with the two-way balance between treated and controls obtained in the covariate balancing propensity score approaches proposed by \citeN{IR:14} and \citeN{zhao2019covariate}.

Entropy balancing was originally proposed by \citeN{hainmueller2012entropy} as a preprocessing method for observational data. A double robustness property was shown by \citeN{zhao2017entropy} for entropy balancing when used as a weighting estimator for the average treatment effect of the treated, ATT. A nonparametric class of calibration estimators, including the entropy balancing estimators was described by \citeN{chan2016globally} and shown to be asymptotically normal and efficient. The version of the estimators studied here, under the assumption of a finite set of covariates balanced, describes a prototypical estimator as it is implemented in practice.
In this paper we show robustness properties for estimation of the average treatment effect, ATE.
Despite the flexibility in the set of assumptions leading to consistency,
in practice it may happen that the number of regressors needed to make these plausible exceeds the number of covariate functions that can be balanced in a given a sample. 
We can include a larger set of covariate functions if we allow for approximate balance \cite{zubizarreta2015stable} rather than imposing exact equality of sample moments. 
However, \citeN{athey2018approximate} claim that the direct use of approximate balancing leads to poor estimates in the case of a high-dimensional regime, and propose a modified version by invoking a regurized linear model for the outcome. 
Strategies for selection of balancing functions, as well as post-selection inference for the resulting estimators, are topics for future research.

\section*{Acknowledgements}
The research was funded by the Swedish Research Council, grant number 2016-00703, and the Marianne and Marcus Wallenberg Foundation. The authors acknowledge the Swedish Childhood Diabetes Study Group. The simulations were run on facilities made avaliable by the High Performance Computing Center North (HPC2N) at Umeå University.

	\addresseshere 	 
	 \bibliographystyle{chicago} 
	 \bibliography{referenser}
	 
\appendix
\section{Proofs}
\begin{lemma}\label{lem:1}
Assumption \ref{as:Ux} implies that $g_t(\lambda) = E[e(X)^t(1-e(X))^{1-t}e^{\lambda^T(U_K(X)- M_K)}]$ has a unique minimizer: exactly one $\tilde{\lambda}_t \in R^K$ satisfies
\[
g^*_t = \inf_{\lambda \in R^K} g_t(\lambda) = g_t(\tilde{\lambda}_t), \quad t = 1,0.
\]
\end{lemma}
\begin{proof} 
Let  $\lambda^*_t = \lambda^*_t (k), k\geq 1,$ be a sequence of vectors such that $\lim_{k \to \infty} g_t(\lambda^*_t) = g_t^*$. 
Without loss of generality, assume $\lambda^*_t /|| \lambda^*_t || \to n^*_t$ as $k \to \infty$, where $n^*_t$ is a normal vector.
If $||\lambda^*_t||$ is bounded, $\lambda^*_t \to \tilde{\lambda}_t$ for some $\tilde{\lambda}_t \in R^K$ with $g_t^* = g_t(\tilde{\lambda}_t)$. 
The assertion follows since $g_t(\lambda)$ is strictly convex, ensuring that the minimizer $\tilde{\lambda}_t$ is unique.

To show that $||\lambda^*_t||$ is bounded, we aim for a contradiction and assume $|| \lambda^*_t || \to \infty$. 
Let $A$ be the event that $\lambda^*_t(U_K(X) -M_K)$ converges in the interval $(-\infty, \infty)$. 
Since $U_K(X)$ has nonsingular covariance matrix (Assumption \ref{as:Ux}), we have $P(A) < 1$ and thus $\lambda^*_t(U_K(X) -M_K)$ tends to $\pm \infty$ with non-zero probability. 
Together with $E[\lambda^*_t(U_K(X) -M_K) ] = 0$, this implies that, with non-zero probability, $\lambda^*_t(U_K(X) -M_K) \to \infty$, leading to $g_t(\lambda^*_t) \to \infty$. This contradicts the definition of $g^*_t$, and hence $|| \lambda^*_t ||$ is bounded and a minimizer of $g_t(\lambda)$ exists.

To conclude, we give a brief argument as to why $g_t(\lambda)$ is strictly convex. For brevity we consider $t=1$. The Hessian matrix $H$ of $g_1(\lambda)$ is given by (row $k$, column $l$)
$$
H_{k,l} = H_{k,l}(\lambda) =  E[e(X)(u_k(X)- m_k)(u_l(X) - m_l)e^{(\lambda)^T(U_K(X) - M_K)}]
$$ 
An estimator of $H_{k,l}$, denoted $\hat{H}_{k,l}$, is obtained by averaging
$$
T_i(u_k(X_i)- m_k)(u_l(X_i) - m_l)e^{(\lambda)^T(U_K(X_i) - M_K)}], \quad i = 1,\ldots,N.
$$
It follows from Assumption \ref{as:Ux} that the matrix $\hat{H}$ is positive-definite with a probability that increases to 1 as $N \to \infty$. Since $E[\hat{H}_{k,l}] = H_{k,l}$, this can be used to verify that $H$ is also positive-definite, which implies strict convexity of $g_1(\lambda)$. \end{proof}\medskip

\begin{lemma} \label{lem:2}
If $Z_1,Z_2,\ldots, Z_N$ are independent copies of a random $n$-vector $Z$ that has a nonsingular covariance matrix, then almost surely the sequence $\{Z_i\}_{i=1}^\infty$ contains $n$ vector observations that are linearly independent.
\end{lemma}
\begin{proof}
Let $A_N$ denote the event that $Z_1,\ldots, Z_N$ contains $n$ linearly independent observations. 
As the covariance matrix of $Z$ is nonsingular, $\lim_{N \to \infty} P(A_N) = 1$, which together with  $A_{N} \subseteq A_{N+1}$ implies $P\left(\cup_{i=1}^\infty A_i \right) =1$. 
Note that $\cup_{i=1}^\infty A_i$ constitute the event of interest, so the assertion follows.
\end{proof} \medskip
\noindent{\bf Proof of Proposition \ref{prop:1}}(i). A key part of the proof is the following property of uniform convergence for approximating the function $g_t(\lambda)$ defined in Lemma \ref{lem:1}: for each compact subset $\Lambda \subset R^K$, 
\begin{equation}\label{unif}
\hat{g}_t(\lambda) = \frac{1}{N} \sum_{i=1}^N T_i e^{(\lambda)^T (U_K(X_i) - \hat{M}_K )} \stackrel{wp1}{\to} g_t(\lambda) \mbox{ uniformly on } \Lambda.
\end{equation}
We briefly describe how to verify (\ref{unif}). Note that
\begin{equation}\label{eq2}
\log \hat{g}_t(\lambda) = 
\log\left( N^{-1} \sum_{i=1}^N T_i e^{(\lambda)^T U_K(X_i)} \right) - (\lambda)^T \hat{M}_K = \log \hat{h}_t(\lambda) - (\lambda)^T \hat{M}_K
\end{equation}
Using the boundedness of $U_K(\cdot)$ (Assumption \ref{as:Ux}) and a uniform convergence result, for example Theorem 6.3 in \cite{SB:13}, it follows that both $\hat{h}_t(\lambda)$ and $(\lambda)^T \hat{M}_K$ converges wp1 uniformly on $\Lambda$, 
which together with the uniform continuity of the logarithmic function implies (\ref{unif}). 

The idea of the proof is the following:
let $Z_{t,i,r}$ denote the indicator function of the event that the dual problem has a unique solution within distance $r>0$ of $\tilde{\lambda}_t$, for sample size $N=i$. 
The aim is to show that almost surely the sequence $\{Z_{t,i}\}_{i \geq 1}$ contains finitely many 0's, i.e,
\begin{equation}
\sum_{i=1}^\infty (1-Z_{t,i,r}) < \infty \quad  \mbox{with probability 1}. \label{zi}
\end{equation}
Since $r>0$ is arbitrary, we obtain from (\ref{zi}) that $\{\hat{\lambda}_t \}$ satisfies
\begin{equation}\label{io}
P(|\hat{\lambda}_t - \tilde{\lambda}_t| > \xi, \; i.o.) = 0,\quad \forall \xi > 0,
\end{equation}
which in turn implies $\hat{\lambda}_t \stackrel{wp1}{\to} \tilde{\lambda}_t$, and the assertion follows. 

To verify property \eqref{zi}, we first show that it holds for $\{W_{t,i,r} \}_{i \geq 1}$, where $W_{t,i,r}$ is the indicator of {\it at least one} (not necessarily a unique) $r$-close solution of the dual problem for sample size $N=i$. 
Given $r>0$, let $\Lambda_{t,r} = \{ \lambda \in R^K : || \lambda - \tilde{\lambda}_t || \leq r \}$ be the ball in $R^K$ with radius $r$ and center at 
$\tilde{\lambda}_t$, and let $\partial \Lambda_{t,r} = \{ \lambda \in R^K : || \lambda - \tilde{\lambda}_t || = r \}$ be its boundary.
Property (\ref{unif}) implies that there exists a (random) number $N^*$ such that (almost surely)
$|\hat{g}_t(\lambda) - g_t(\lambda)| < r/2, \lambda \in \partial \Lambda_{t,r}$, and $|\hat{g}_t(\tilde{\lambda}_t) - g_t(\tilde{\lambda}_t)| < r/2$ for $N \geq N^*$. 
Since $\tilde{\lambda}_t$ minimizes $g_t(\lambda)$, the convex function $\hat{g}_t(\lambda)$ has a local minima in $ \Lambda_{t,r}$ for $N\geq N^*$, and hence the gradient $\nabla_\lambda \hat{g}_t(\hat{\lambda}_t) = 0^K$ for some $\hat{\lambda}_t$, i.e.,
\begin{equation}\label{eq:lemma2}
\sum_{\{i:T_i=t \}}  e^{(\hat{\lambda}_t)^T U_K(X_i) } [u_k(X_i) - \hat{m}_k] = 0, \quad k = 1, \ldots, K, 
\end{equation}
and so the dual problem has at least one soluton when $N\geq N^*$, implying $W_{t,i,r} = 1$, for $i \geq N^*$, and thus property (\ref{zi}) holds for $\{W_{t,i,r}\}_{i \geq 1}$.

Next we examine under which conditions a given solution is unique.
The Hessian matrix of $\hat{g}_{t}(\lambda)$, denoted by $H_t$, has elements (row $l$, column $m$)
\begin{equation}\label{H}
H_{t}(l,m) = \frac{1}{N}\sum_{\{i.T_i=t\}} (u_l(X_i) - \hat{m}_l)(u_m(X_i) - \hat{m}_m) e^{(\hat{\lambda}_t)^T (U_K(X_i) - \hat{M}_K)}.
\end{equation}
By \eqref{H}, if the observed covariate vectors $\{ X_i : T_i=t \}$ in group $t$ have full rank $K$, the Hessian $H_t$ is positive definite. 
This ensures strict convexity of $\hat{g}_t(\lambda)$, i.e., that at most one minimizer of $\hat{g}_t(\lambda)$ is possible. 
By Assumption \ref{as:Ux}, the vector $TU_K(X)$ has a nonsingular covariance matrix, and so by Lemma \ref{lem:2}, almost surely a finite number of the 1's in $\{W_{t,i,r}\}_{i \geq 1}$ correspond to multiple solutions, implying that almost surely $\{W_{t,i,r}\}_{i \geq 1}$ and $\{Z_{t,i,r}\}_{i \geq 1}$ differ on a finite number of positions, and thus (\ref{zi}) holds. This completes the proof. \qed  \medskip

\noindent{\bf Proof of Proposition \ref{prop:1}}(ii).
An implication of Assumption \ref{as:Ux} is that the $(K+1)$-vectors $(T,TU_K(X))$ and $(1-T,(1-T) U_K(X))$ have nonsingular covariance matrices. 
Thus, Lemma \ref{lem:2} yields that (almost surely) the sequence $\{\hat{\kappa}_t \}$ after a finite number of steps is given by the unique dual solution $\hat{\kappa}_t = (N^{-1}A_t^T A_t)^{-1} \hat{M}^*_{K}, t =1,0$. By combining this result with the strong law of large numbers, the assertion follows. \qed \medskip

\noindent{\bf Proof of Theorem \ref{th:1}}. 
We have
\begin{equation}
\hat{\beta}_{\mbox{\scriptsize KL},t} = \frac{1}{N} \sum_{\{i:T_i=t\}}  w_{t,i}^{\mbox{\scriptsize KL}}   Y_i =
\frac{N^{-1}\sum_{\{i:T_i=t\}}  e^{(\hat{\lambda}_t)^T U_K(X_i)}   Y_i}
{N^{-1}\sum_{\{i:T_i=t\}}  e^{(\hat{\lambda}_t)^T U_K(X_i)}},
\end{equation}
which is a fraction of two averages, each containing the dual parameter $\hat{\lambda}_t$.
The limit  $\hat{\lambda}_t \stackrel{wp1}{\to} \tilde{\lambda}_t$ in Proposition \ref{prop:1},
together with a law of large numbers for averages with estimated parameters \cite[Th.\ 7.3]{SB:13}, leads to
\begin{equation}
\hat{\beta}_{\mbox{\scriptsize KL},t} \stackrel{P}{\to}  \frac{E[T^t (1-T)^{1-t}e^{(\tilde{\lambda}_t )^T U_K(X)}Y(t) ]}{E[T^t (1-T)^{1-t}e^{(\tilde{\lambda}_t )^T U_K(X)}]}  = E[e(X)^t(1-e(X))^{1-t} \tilde{w}^{\mbox{\scriptsize KL}}_{t}(X) \beta_1(X)],
\end{equation}
so the assertion follows. The same argument holds for the quadratic estimator. \qed \medskip

\noindent{\bf Proof of Theorem \ref{th:norm}}.
 We have
\begin{equation}\label{c0}
\esto{\kl,1} =  \frac{1}{N} \sum_{\{i:T_i=1\}}  w_{1,i}^{\mbox{\scriptsize KL}} Y_i =  \frac{1}{N} \sum_{\{i:T_i=1\}}  w_{1,i}^{\mbox{\scriptsize KL}} \beta_1(X_i) + 
\frac{1}{N} \sum_{\{i:T_i=1\}} w^{\mbox{\scriptsize KL}}_{1,i} \epsilon_{1,i}, \\
\end{equation}
where $\epsilon_{1,i} := Y_i(1) - \beta_1(X_i)$.
We use the decomposition
\begin{equation} \label{c1} 
\frac{1}{N} \sum_{\{i:T_i=1\}} w^{\mbox{\scriptsize KL}}_{1,i} \epsilon_{1,i} 
= \frac{1}{N}\sum_{\{i:T_i=1\}} \tilde{w}^{\mbox{\scriptsize KL}}_{1}(X_i) \epsilon_{1,i} + R_{1,N} R_{2,N} + R_{3,N} R_{4,N},
\end{equation} 
where
\[
\begin{array}{l}
R_{1,N}  = \frac{1}{N^{-1}\sum_{T_i=1} e^{{(\hat{\lambda}_1})^T U_K(X_i)} } - \frac{1}{E[e(X)e^{(\tilde{\lambda}_1)^T U_K(X)}]},  \\
R_{2,N}  =  \frac{1}{N} \sum_{\{i:T_i=1\}} e^{(\tilde{\lambda}_1)^T U_K(X_i)} \epsilon_{1,i},  \\ 
R_{3,N}   =  \frac{1}{N^{-1}\sum_{T_i=1} e^{{(\hat{\lambda}_1})^T U_K(X_i)} }, \\
R_{4,N}  =  \frac{1}{N} \sum_{\{i:T_i=1\}} \left[ e^{(\hat{\lambda}_1)^T U_K(X_i)} - e^{(\tilde{\lambda}_1)^T U_K(X_i)} \right] \epsilon_{1,i}. 
\end{array}
\]
From results for averages with estimated parameters, as in the proof of Theorem \ref{th:1}, 
\begin{align}
R_{1,N} &  \stackrel{P}{\to} 0 \mbox { as } N \to \infty, \label{c2} \\
R_{3,N} &  \stackrel{P}{\to} c \mbox { as } N \to \infty, \mbox{ for some $c>0$}. \label{c3}
\end{align}
The term $R_{2,N}$ is an average of independent terms with mean zero (from Assumption \ref{nuc}) and finite variance (since $V[Y(1)] < \infty$), 
so the central limit theorem yields
\begin{equation} \label{c4}
\sqrt{N} R_{2,N} \stackrel{D}{\to} N(0, Var[Te^{(\tilde{\lambda}_1)^T U_K(X_i)} \epsilon_{1,i}]) \mbox{ as } N \to \infty.
\end{equation}
By Assumption \ref{nuc}, $R_{4,N}$ consists of uncorrelated terms with zero mean, and therefore 
\begin{equation}\label{c6}
E\left[ (\sqrt{N}R_{4,N})^2 \right]  = E\left[ \left( e^{(\hat{\lambda}_1)^T U_K(X_i)} - e^{(\tilde{\lambda}_1)^T U_K(X_i)} \right )^2 \epsilon_{1,i}^2 \right] \to 0  \mbox{ as } N \to \infty,
\end{equation}
where the limit is a consequence of the dominated convergence theorem: (i) Proposition \ref{prop:1} yields that the sequence inside the expectation almost surely tends to zero, and (ii) the boundedness of $U_K(X)$ (Assumption \ref{as:Ux}) and $\hat{\lambda}_1$ (Proposition \ref{prop:1}), combined with $E[Y(1)^2] <~\infty$, imply that the sequence is bounded by a random variable with finite expectation.

Now, \eqref{c2} and \eqref{c4} yield $\sqrt{N} R_{1,N} R_{2,N} \stackrel{P}{\to} 0$, and from \eqref{c3} and \eqref{c6} we get $\sqrt{N} R_{3,N} R_{4,N} \stackrel{P}{\to} 0$. 
Moreover, the results on asymptotic existence of the dual solutions derived in the proof of Proposition \ref{prop:1}, together with the assumed linearity of $\beta_1(X)$ and $\beta_0(X)$, implies
$\sqrt{N} W_N \stackrel{wp1}{\to} 0$, where
\begin{equation}
 W_N =  \frac{1}{N} \sum_{\{i:T_i=1\}}  w_{1,i}^{\mbox{\scriptsize KL}} \beta_1(X_i) - \frac{1}{N} \sum_{t=1} ^N \beta_t(X_i).
\end{equation}
These limit results in conjunction with \eqref{c1} entail that $\hat{\beta}_{\mbox{\scriptsize KL},1}$ can be partitioned as
\begin{equation}
\hat{\beta}_{\mbox{\scriptsize KL},1} = \frac{1}{N} \sum_{i=1}^N \beta_1(X_i) + \frac{1}{N} \sum_{\{i:T_i=1\}} \tilde{w}^{\mbox{\scriptsize KL}}_{1}(X_i) \epsilon_{1,i}+ R_N,
\end{equation}
and where $\sqrt{N}R_N \stackrel{P}{\to} 0$.
A corresponding result can be obtained for $\hat{\beta}_{\mbox{\scriptsize KL},0}$, and so the claimed asymptotic normality of 
$\hat{\beta}_{\mbox{\scriptsize KL}} = \hat{\beta}_{\mbox{\scriptsize KL},1}-\hat{\beta}_{\mbox{\scriptsize KL},0} $ follows from conventional asymptotic results, i.e., the central limit theorem and Slutsky's theorem.
Similar arguments hold for the quadratic estimator $\hat{\beta}_{\mbox{\scriptsize QR}}$. 
This completes the proof. \qed

\section{Small bias for a zero treatment effect}
The aim here is to address why the estimators $\est{(1)}{\kl}$ and $\est{(1)}{\qd}$ have negligible bias for a zero treatment effect in designs B and C. This can not be explained by Corollary 2(i) since the conditional outcome $\beta(X) = E[Y(t) | X], t=1,0,$ is non-linear in the balanced functions $u_k(X) = X_k, k=1,\ldots, 6$, thus violating the assumptions. We present two other scenarios in which entropy balancing leads to small estimation error, and then argue that the considered designs can be described as hybrids of these scenarios, thereby explaining the small bias. Detailed arguments are given for Kullback-Leibler estimator $\est{(1)}{\kl}$. For the quadratic estimator $\est{(1)}{\qd}$, parts of the arguments are valid, such as scenario 1, but further investigations are needed to explain its small bias.

In general, the assumptions for consistency of $\esto{\kl} = \esto{\kl,1} - \esto{\kl,0}$ in Corollary 2 ensure that the conditional means $\beta_1(X)$ and $\beta_0(X)$ are three-way balanced, implying that each of the components $\esto{\kl,1}$ and $\esto{\kl,0}$ are consistent as estimators of the corresponding parameters $\beta_1 = E[Y(1)]$ and $\beta_0 = E[Y(0)]$. However, for a zero treatment effect, the asymptotic estimation error (Corollary 1) is given by
\begin{equation*}
\hat{\beta}_{KL} - 0 \stackrel{P}{\to} E[e(X) \tilde{w}^{KL}_1 (X)\beta(X)] - E[(1-e(X)) \tilde{w}^{KL}_0 (X)\beta(X)]  \mbox{  as  } N \to \infty,
\end{equation*}
wherefore it is sufficient for consistency of $\esto{\kl}$ that the common conditional mean $\beta(X)$ is two-way balanced between the treatment and control groups, i.e., that
\begin{equation}\label{twoway}
E[e(X) \tilde{w}^{KL}_1 (X)\beta(X)] = E[(1-e(X)) \tilde{w}^{KL}_0 (X)\beta(X)].
\end{equation}
Assuming that the asymptotic properties can explain our observed simulated results,  
we argue that the small bias is expected by showing that two-way balance \eqref{twoway} holds approximately. 

\subsection*{Scenario 1. A symmetric design} 
Assume that $\esto{\kl}$  balances a function $U_K(X)$ that is symmetric in the sense that $U_K(X)$ and $-U_K(X)$ have the same distribution.
Let the treatment variable be given by $T = I(\eta_K^T U_K(X) + \epsilon > 0), \eta_K \in R^K $, where $\epsilon$ and $-\epsilon$ have the same distribution. Then it is straightforward to verify that
\begin{equation}\label{mirror}
P(U_K(X) \in A | T = 1) = P(-U_K(X) \in A | T = 0), \quad A \subset R^K,    
\end{equation}
meaning that the (conditional) distribution of $U_K(X)$ given $T=1$ is the same as the distribution of $-U_K(X)$ given $T = 0$. 
This 'mirroring' property together with Proposition~\ref{prop:1}(i) implies that the asymptotic dual parameters fulfill $\tilde{\lambda}_1 = -\tilde{\lambda}_0$, which in turn implies that the limiting weights satisfy
\begin{equation}\label{weights}
	\tilde{w}_1^{KL} (X) = \tilde{w}_0^{KL} (-X)
\end{equation}
Now, using properties \eqref{mirror} and \eqref{weights} and $P(T=1) = 1/2$, we get 
\begin{align*}
E[e(X) \tilde{w}^{KL}_1 (X)\beta(X)] & = E[\tilde{w}^{KL}_0 (-X) \beta(X) |T=1] P(T=1) \\
& = E[\tilde{w}^{KL}_0 (X) \beta(-X) |T=0] P(T=0) \\
& =E[(1-e(X)) \tilde{w}^{KL}_0 (X)\beta(-X)]
\end{align*}
Thus, if we additionaly assume that $\beta(X) = \beta(-X)$, it follows that \eqref{twoway} holds and $\esto{\kl}$ is consistent. 
A similar argument can be used to show that $\esto{\qd}$ is also consistent in this scenario. \medskip

\noindent
{\it Example.}
Let $X \sim Unif(-4,4)$ and $Y(t) = X^2 + \epsilon_t$, where $\epsilon_t \sim N(0,1)$. Then $\esto{\kl}$, which balances the univariate function $U_1(X) = X$, is consistent, $\esto{\kl} = \esto{\kl,1} - \esto{\kl,0}  \stackrel{P}{\to} 0$, but for the component estimators we have $\esto{\kl,1},\esto{\kl,0} \approx 1.75$, which deviates substantially from the corresponding true value $E[X^2] = 16/3 \approx 5.33$.  \medskip

\subsection*{Scenario 2. Exponential tilting of gamma distributions} The second scenario is for the covariate $X_5 \sim \chi^2(1)$, which falls outside of scenario 1 because of its asymmetric distribution and the term $\sqrt{X_5}$ in $\beta(X)$ in design B. Heuristically, the reason why $X_5$ does not contribute with bias is that its density has a shape that is "tilted back" to a $\chi^2(1)$-distribution by the exponential KL weights.

First we give a general description of this scenario. Let $X$ be gamma distributed with shape parameter $k$ and scale parameter $\theta$, i.e., the density is
\[
f(x|k,\theta) = \frac{1}{\Gamma(k) \theta^k} x^{k-1}  e^{-x/\theta}, \quad x >0.
\]
Now we make the assumption that the conditional distribution of $X$ in group $t$ can be closely approximated with a gamma distribution with the same shape parameter $k$ but with a group-dependent scale parameter $\theta_t$. 
Then, if $\hat{\beta}_{KL}$ is obtained by balancing $X$ only, we claim that we get (asymptotic) three-way balance
\begin{equation}\label{threewaytilt}
E[T\tilde{w}^{KL}_1 (X) g(X) ] =  E[(1-T)\tilde{w}^{KL}_0 (X) g(X) ] = E[g(X)]
\end{equation}
for (essentially) any function $g(X)$. We argue briefly why \eqref{threewaytilt} holds. Consider the asymptotic weight for the treatment group, 
\begin{equation*}
\tilde{w}^{KL}_1 (X) =   \frac{e^{\tilde{\lambda}_{1,1} X} }{E[T e^{\tilde{\lambda}_{1,1} X} ]} =  e^{\tilde{\lambda}_{1,0} + \tilde{\lambda}_{1,1} X}, 
\end{equation*}
where $\tilde{\lambda}_{1,1}$ is obtained from the balance equation
\begin{equation*}
E[T  e^{\tilde{\lambda}_{1,1} X}(X - E[X])]   = 0.
\end{equation*}
Since the solution $\tilde{\lambda}_{1,1}$ is unique and the conditional density of $X$ given $T=1$ is  $f(x|k,\theta_1)$, it must hold that $\tilde{\lambda}_{1,1} - 1/\theta_1 = -1/\theta$ and $E[T  e^{\tilde{\lambda}_{1,1} X}] = \sqrt{\theta/\theta_1} P(T=1)$, which altogether implies  $E[T\tilde{w}^{KL}_1 (X) g(X) ] = E[g(X)]$, and similarly for the control group, implying that \eqref{threewaytilt} holds.

\begin{figure}[htbp]
\begin{center}
\includegraphics[width=\textwidth]{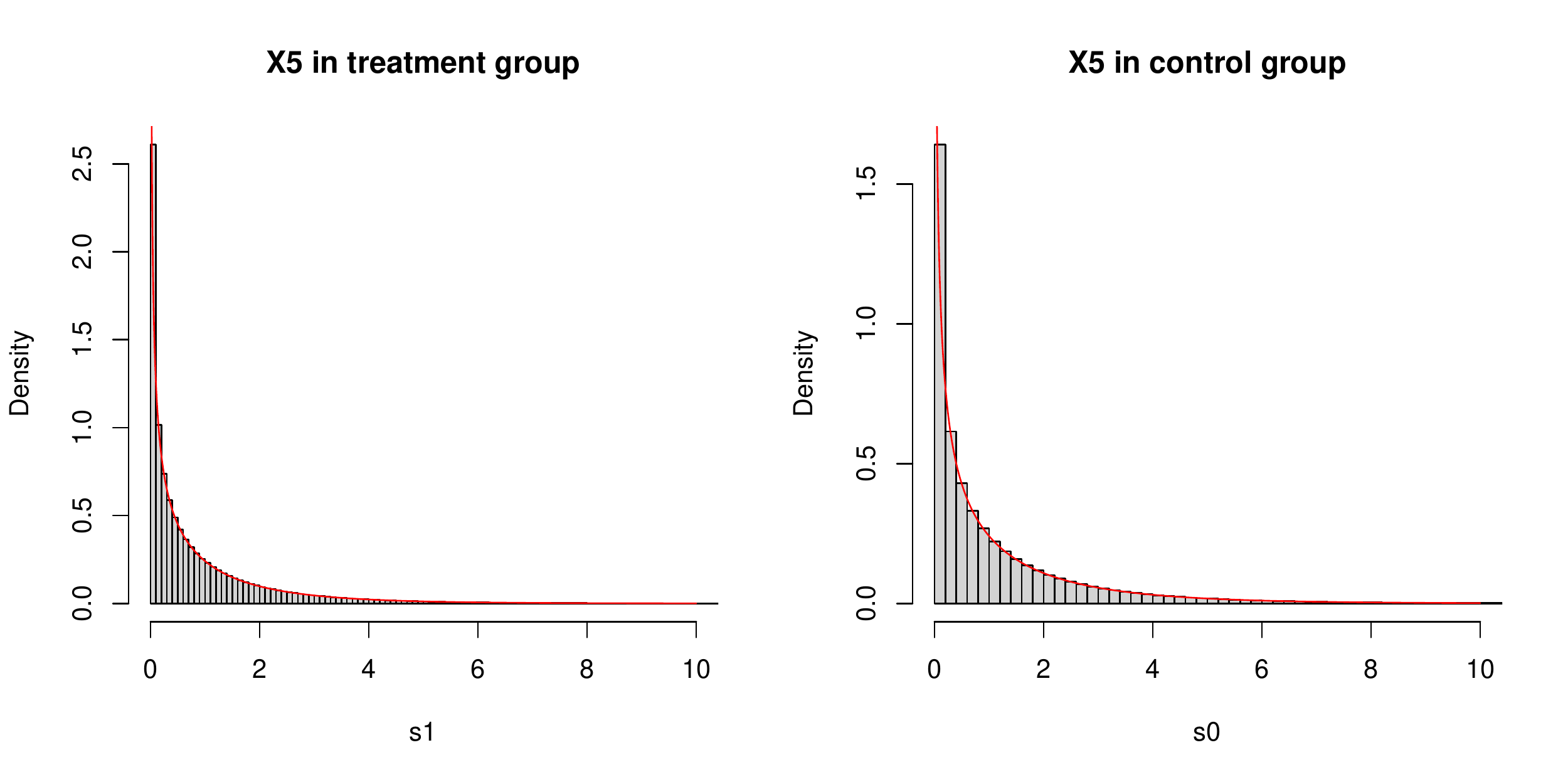}
\caption{Density estimation of $X_5$ in each group. The red lines depict gamma densities with shape $k=1/2$ and group-dependent scale parameters 
$\hat{\theta}_1$ and $\hat{\theta}_0$, which are the maximum likelihood estimates in respective group.}
\label{Fig1}
\end{center}
\end{figure}

Now, the covariate $X_5$ in our simulations is $\chi^2(1)$-distributed, i.e.\ gamma distributed with density  $f(x|1/2,2)$, and we have that (see Figure \ref{Fig1}) the conditional distribution in respective group is also gamma with shape $k=1/2$. Following the discussion above, the KL exponential weights that balance $X_5$ also three-way balance arbitrary functions $g(X_5)$, as in \eqref{threewaytilt}. Note that we can examine the anticipated relation $\tilde{\lambda}_{1,1} - 1/\theta_1 = \tilde{\lambda}_{0,1} - 1/\theta_0  = -1/2$ using the maximum-likelihood estimates $\hat{\theta}_1$ and $\hat{\theta}_0$ and dual parameters $\hat{\lambda}_{1,0}$ and $\hat{\lambda}_{1,1}$ from a large simulated dataset. For example, from a sample of size $N=10,000$ we obtained $\hat{\lambda}_{1,1} - 1/\hat{\theta}_1 \approx -0.496$ and $\hat{\lambda}_{1,0} - 1/\hat{\theta}_0 \approx -0.494$.

\subsection*{Hybrids of scenarios 1 and 2}
In the following we show that the KL weights (approximately) two-way balance all non-linear and interaction terms, $\sqrt{X_5}$, $X_1 X_2$ etc., that are the constituents of the conditional mean $\beta(X)$ in designs B and C. 
 
A key assumption (A1) here is that $X_5$ and $X_{-5} = (X_1,X_2,X_3,X_4,X_6)$ are (conditionally) independent in the treatment and control groups. This allows us to merge scenarios 1 and 2. Clearly A2 is an approximate statement, and it might seem strong and difficult to validate, but we claim it is plausible given the relatively high variance of the overlap term $\epsilon$.  

Henceforth we replace $X_6$ as a balancing function with its mean centered version defined by $X^*_6 = X_6 - 0.5$, which is only a matter of notation since, due to the normalization constraint for the weights, the estimator is left unchanged. The treatment variable can be written as
\[
T = I( X_1 + 2X_2 -2 X_3 - X_4 + X_6^* - 0.5 (X_5-1)  + \epsilon > 0)
\]
Note that $X_{-5} = (X_1,X_2,X_3,X_4,X_6^*)$ obeys the 'mirroring' property of the balancing vector in scenario~1, provided that the distribution of the zero mean variable $- 0.5 (X_5-1)  + \epsilon$ is (approximately) symmetric around zero. This assumption (A2) holds with high precision, as is straightforward to verify using simulations. 

Assumptions A1 and A2 allow us to use asymptotic properties of the weights that are valid when $X_5$ and $X_{-5}$ are balanced separately, i.e., scenarios 1 and 2. This is used next to verify that $\est{(1)}{\kl}$ (approximately) two-way balances each term in $\beta(X)$ in designs B and C. A consequence of A1 is that the dual limit vector for $X_{-5} = (X_1,X_2,X_3,X_4,X_6)$, denoted here by $\tilde{\lambda}_{t,-5}$, does not depend on wether $X_5$ is balanced or not, and vice versa. This becomes visible if A1 is used to decompose the balance equations in Proposition \ref{prop:1}(i). 
Now, using A2, we get that property \eqref{twoway} holds for the quadratic terms $X_1^2$ and $X_2^2$ in design C since
\begin{align*}
E[T\tilde{w}_1^{\mbox{\scriptsize KL}}(X) X_1^2] & = \frac{E[T e^{(\tilde{\lambda}_1)^T U_6(X)} X_1^2]}{E[T e^{(\tilde{\lambda}_1)^T U_6(X)}]} 
= \frac{E[e^{\tilde{\lambda}_{1,5} X_5}|T=1] E[e^{\tilde{\lambda}_{1,-5} X_{-5}} X_1^2|T=1]}{E[e^{\tilde{\lambda}_{1,5} X_5}|T=1] E[e^{\tilde{\lambda}_{1,-5} X_{-5}}|T=1]}  \\
& = \frac{E[e^{\tilde{\lambda}_{0,5} X_5}|T=0] E[e^{\tilde{\lambda}_{0,-5} X_{-5}} (-X_1)^2|T=0]}{E[e^{\tilde{\lambda}_{0,5} X_5}|T=0] E[e^{\tilde{\lambda}_{0,-5} X_{-5}}|T=0]} \\ 
&= E[(1-T)\tilde{w}_0^{\mbox{\scriptsize KL}}(X) X_1^2] 
\end{align*}
Similarly, for a function $g(X_5)$, e.g., $\sqrt{X_5}$ in design B and $X_5^2$ in design C, we have
\begin{align*}
E[T\tilde{w}_1^{\mbox{\scriptsize KL}}(X) g(X_5)] & = \frac{E[e^{\tilde{\lambda}_{0,5} X_5} g(X_5)|T=0] E[e^{\tilde{\lambda}_{0,-5} X_{-5}} |T=0]}{E[e^{\tilde{\lambda}_{0,5} X_5}|T=0] E[e^{\tilde{\lambda}_{0,-5} X_{-5}}|T=0]} = E[g(X_5)] \\
& =  E[(1-T)\tilde{w}_0^{\mbox{\scriptsize KL}}(X) g(X_5)].
\end{align*}
Two-way balance of the interaction terms $X_1 X_5$ and $X_2 X_5$ can be shown in a similar way. This completes the argumentation for the zero treatment effect designs. 

As a final remark, since we have shown that the Kullback-Leibler weights three-way balance all functions of $X_5$, we do not expect the term $\sqrt{X_5}$ in design B to contribute to the bias of $\est{(1)}{\kl}$ for the heterogeneous treatment effect. This is a possible explanation why $\est{(1)}{\kl}$  has a smaller bias than $\hat{\beta}^{(1)}_{\mbox{\scriptsize \qd}}$ in this design.

\section{Covariate description for data example: Acute complications of Type 1 Diabetes Mellitus}
In Table \ref{desc} we display summary data for the 22 pre-treatment variables in the analysis of the data example.  The parents' socioeconomic variables are measured one year prior to receiving the grade. For the IPW-LOGIT estimator a first order logistic regression model is fitted. The entropy balancing estimators use the sample means of each variable as balance constraints. 

\begin{table}%[ht!]  
\caption{Statistics for group proportions (\%) and mean values for the treated (low grades) and control-subgroups (higher grades). 
Categories for the education variable: Compulsory Schooling (CS), Upper Secondary Education (SE) and Higher Education (HE). } 
\centering
\begin{tabular}{ l  c  c }
	 \hline
	Variable  										    & Low grades 	& Higher grades  \\ 
	 \hline
	Mean number of hospitalization 	                    &	 1.5			& 	0.6		\\
	with acute complications && \\
	Female gender  						    &  	37.1\%			& 	51.2\%  \\
	Mean age at onset of T1DM				&	9.1 years   		& 	9.0 years  \\
	Mother's mean age at delivery 			&	27.8 years		&	29.0 years \\
	Father's mean age at delivery 			&	30.9 years		&	31.8 years \\
	Social benefits in the family 	        &				9.9\% 			& 	3.0\%	\\
	Single mother status				    &			33.4\%   			& 	19.3\%  \\
	Education level	(CS/SE/HE)			&&	\\
	\, \, - mother 						&  25.9\% /57.9\% /16.2\%    &	10.9\% /50.6\% /38.5\%   \\
	\, \, - father						&  30.6\% /57.8\% /11.6\%  &	 18.2\% /48.1\% /33.7\%  \\ 
	Mean salary income  	&	\\
	\, \, - mother						&	128 500 SEK			&	171 400 SEK	\\
	\, \, - father						&	201 400 SEK			&	273 000 SEK  \\
	Sick pay			&\\
	\, \, - mother						& 23.8\%				&	16.9\% \\
	\, \, - father						& 17.6\%				&	11.6\% \\
	Unemployment benefits &	\\
	\, \, - mother						& 16.2\%				&	10.4\% \\
	\, \, - father					& 12.8\%				&	6.5\% \\
	Benefits for labour market policy measures &\\
	\, \, - mother						& 5.9\%				&	 3.6\%     \\
	\, \, - father						& 5.1\%				&	2.8\%	\\
	Early retirement and sick benefits	&\\
	\, \, - mother							& 12.4\%				&	6.3\% \\
	\, \, - father							& 7.3\%				&	4.2\% \\
	Retirement pension				&	\\
	\, \, - mother							& 1.7\%				&	0.7\% \\
	\, \, - father							& 2.9\%				&	2.3\% \\
	Occupational injury annuity 	&	\\ %(YrkArbTyp)
	\, \, - mother							& 1.7\%				&    0.7\% \\
	\, \, - father					& 3.1\%				&    1.5\% \\	
	 \hline
\end{tabular}
\label{desc}
\end{table}

\label{lastpage}
\end{document}